\documentclass[traditabstract]{aa}  
\usepackage{graphicx}
\usepackage{amsmath}

\newcommand{\Msol}{ {\cal M}_{\sun}}
\newcommand{\M}{ {\cal M}}

\newcommand{\Ms}{ {\cal M_\star}}
\newcommand{\SFR} { {\cal SFR}}
\newcommand{\MF} { {{\cal GSMF}_{SF}} }

\begin{document}
   \title{Evolution of the specific star formation rate function at $z<1.4$ \\ Dissecting the mass-$\SFR$ plane in COSMOS and GOODS}

  \titlerunning{Evolution of the specific Star Formation Rate Function at $z<1.4$}

   \author{
O. Ilbert \inst{1}
\and S.~Arnouts \inst{1}
\and E.~Le Floc'h \inst{2}
\and H.~Aussel \inst{3}
\and M.~Bethermin \inst{4}
\and P.~Capak \inst{5}
\and B.-C. Hsieh \inst{6}
\and M.~Kajisawa \inst{7,8}
\and A.~Karim \inst{9}
\and O.~Le F\`evre \inst{1}
\and N.~Lee \inst{10}
\and S.~Lilly \inst{11}
\and H.~J.~McCracken \inst{12}
\and L.~Michel-Dansac \inst{13}
\and T.~Moutard \inst{1}
\and M.~A. Renzini \inst{14}
\and M.~Salvato \inst{15}
\and D.~B. Sanders \inst{9}
\and N.~Scoville \inst{16}
\and K.~Sheth \inst{17}
\and J.D.~Silverman \inst{18}
\and V.~Smol{\v c}i{\'c} \inst{19}
\and Y.~Taniguchi \inst{7}
\and L.~Tresse \inst{1}
         }
\institute{
Aix Marseille Universit\'e, CNRS, LAM (Laboratoire d'Astrophysique de Marseille) UMR 7326, 13388, Marseille, France,    \email{olivier.ilbert@lam.fr}
\and
AIM Unit\'e Mixte de Recherche CEA  CNRS Universit\'e Paris VII UMR n158, France
\and
Laboratoire AIM, CEA/DSM/IRFU, CNRS, Universit\'e Paris-Diderot, 91190, Gif, France
\and
European Southern Observatory, Karl-Schwarzschild-Str. 2, 85748 Garching, Germany
\and
Spitzer Science Center, California Institute of Technology, Pasadena, CA 91125, USA
\and
Institute of Astronomy and Astrophysics, Academia Sinica, P.O. Box 23-141, Taipei 10617, Taiwan
\and
Research Center for Space and Cosmic Evolution, Ehime University, 2-5 Bunkyo-cho, Matsuyama 790-8577, Japan 
\and
Physics Department, Graduate School of Science \& Engineering, Ehime University, 2-5 Bunkyo-cho, Matsuyama 790-8577, Japan 
\and
Argelander-Institut f\"ur Astronomie, Universit\"at Bonn, Auf dem H\"ugel 71, D-53121 Bonn, Germany
\and
Institute for Astronomy, 2680 Woodlawn Dr., University of Hawaii, Honolulu, Hawaii, 96822
\and
Department of Physics, ETH Zurich, CH-8093 Zurich, Switzerland
\and
Institut d'Astrophysique de Paris, UMR7095 CNRS, Universit\'e Pierre et Marie Curie, 98 bis Boulevard Arago, 75014 Paris, France
\and
Centre de Recherche Astrophysique de Lyon, Universit\'e de Lyon, Universit\'e Lyon 1, Observatoire de Lyon, France.
\and
Dipartimento di Astronomia, Universita di Padova, vicolo dell'Osservatorio 2, I-35122 Padua, Italy
\and
Max-Planck-Institut f\"ur Extraterrestrische Physik, Postfach 1312, D-85741, Garching bei M\"unchen, Germany
\and
California Institute of Technology, MC 105-24, 1200 East California Boulevard, Pasadena, CA 91125
\and
National Radio Astronomy Observatory, 520 Edgemont Road, Charlottesville, VA 22903, USA
\and
Kavli Institute for the Physics and Mathematics of the Universe, Todai Institutes for Advanced Study, the University of Tokyo, Kashiwa, Japan 277-8583 (Kavli IPMU, WPI)
\and
University of Zagreb, Physics Department, Bijeni\v{c}ka cesta 32, 10002 Zagreb, Croatia
}

   \date{Received ... ; accepted ...}

   \abstract{The relation between the stellar mass ($\Ms$) and the
     star formation rate ($\SFR$) characterizes how the instantaneous
     star formation is determined by the galaxy past star formation
     history and by the growth of the dark matter structures. We
     deconstruct the $\Ms-\SFR$ plane by measuring the specific $\SFR$
     functions in several stellar mass bins from $z=0.2$ out to
     $z=1.4$ (specific $\SFR=\SFR/\Ms$, noted $s\SFR$). Our analysis
     is primary based on a 24$\mu m$ selected catalogue combining the
     COSMOS and GOODS surveys. We estimate the $\SFR$ by combining
     mid- and far-infrared data for $20500$ galaxies. The $s\SFR$
     functions are derived in four stellar mass bins within the range
     $9.5<log(\Ms/\Msol)<11.5$. First, we demonstrate the importance
     of taking into account selection effects when studying the
     $\Ms-\SFR$ relation. Secondly, we find a mass-dependent evolution
     of the median $s\SFR$ with redshift varying as $s\SFR \propto
     (1+z)^{b}$, with $b$ increasing from $b=2.88^{\pm 0.12}$ to
     $b=3.78^{\pm 0.60}$ between $\Ms=10^{9.75}\Msol$ and
     $\Ms=10^{11.1}\Msol$, respectively. At low masses, this evolution
     is consistent with the cosmological accretion rate and
     predictions from semi-analytical models (SAM). This agreement
     breaks down for more massive galaxies showing the need for a more
     comprehensive description of the star formation history in
     massive galaxies. Third, we obtain that the shape of the $s\SFR$
     function is invariant with time at $z<1.4$ but depends on the
     mass. We observe a broadening of the $s\SFR$ function ranging
     from 0.28 dex at $\Ms=10^{9.75}\Msol$ to 0.46 dex at
     $\Ms=10^{11.1}\Msol$. Such increase in the intrinsic scatter of
     the $\Ms-\SFR$ relation suggests an increasing diversity of SFHs
     as the stellar mass increases. Finally, we find a gradual decline
     of the $s\SFR$ with stellar mass as $log_{10}(s\SFR) \propto
     -0.17\Ms$. We discuss the numerous physical processes, as gas
     exhaustion in hot gas halos or secular evolution, which can
     gradually reduce the $s\SFR$ and increase the SFH diversity.

     \keywords{Galaxies: distances and redshifts -- Galaxies:
       evolution -- Galaxies: formation -- Galaxies: star formation --
       Galaxies: stellar content}}

   \maketitle
%

\section{Introduction}

Numerous observational results show a tight relationship between the
stellar mass ($\Ms$) and the star formation rate ($\SFR$) of
star-forming galaxies (e.g. Noeske et al. 2007a, Elbaz et al. 2007,
Daddi et al. 2007, Peng et al. 2010, Karim et al. 2011, Elbaz et
al. 2011). The star-forming galaxies are distributed in the
$\Ms$-$\SFR$ plane along what is commonly called the ``star-forming
Main Sequence''. If we do not consider quiescent galaxies, the
existence of such a $\Ms$-$\SFR$ relation implies that the galaxies
that are currently the most star-forming were also the most
star-forming in their past history. Star-forming galaxies are
scattered around this relation as expected from the stochasticity in
their individual star formation histories (SFHs) (e.g. Hopkins et
al. 2014, Dom{\'{\i}}nguez et al. 2014) and from the variety of
possible SFHs. Extreme events like mergers could decouple the
instantaneous $\SFR$ from the past star formation history and create
outliers to the $\Ms$-$\SFR$ relation, which is one definition of
starbursts (e.g. Rodighiero et al. 2011).

While the shape and the scatter of the $\Ms$-$\SFR$ relation already
provide deep insights into the galaxy assembly process, its evolution
along cosmic time is also of a great interest. Noeske et al. (2007a),
Daddi et al. (2007) and Elbaz et al. (2007) find that the $\Ms$-$\SFR$
relation scales with cosmic time, such that the $\SFR$ increases with
redshift at a given stellar mass. This evolution is also seen as an
increase in the specific $\SFR$ (hereafter $s\SFR=\SFR/\Ms$) with
redshift at a given stellar mass. There is a growing consensus that
the $s\SFR$ evolution is deeply linked to the hierarchical growth of
dark matter structures (e.g. Bouch\'e et al. 2010, Lilly et
al. 2013). Assuming that galaxies are fed by fresh gas at a constant
fraction of the averaged cosmological accretion rate, the $s\SFR$
should evolve as the specific dark matter increase rate (hereafter
$sMIR_{DM}$) defined as $\dot{M}_{H}/M_H$ with $M_H$ the mass of the
dark matter halos (Lilly et al. 2013). However, the galaxies could be
fed more efficiently in fresh gas at high redshift than in the local
Universe since cold accretion occurs mainly at $z>2$ (e.g. Dekel et
al. 2009). Therefore, having an accurate characterization of the
$s\SFR$ evolution with redshift is crucial in order to link the galaxy
stellar mass assembly with the growth of the dark matter structures.

Below $z<1-1.5$, the $s\SFR$ is relatively well measured using robust
infrared data (Noeske et al. 2007a, Elbaz et al. 2007, Elbaz et
al. 2011) and radio data (Karim et al. 2011). The $s\SFR$ increases
steadily from the present day to $z \sim 2$ (e.g. Daddi et al. 2007,
Karim et al. 2011) and its evolution is usually parametrized as $s\SFR
\propto (1+z)^{b}$. The values of $b$ in the literature cover the full
range between 2.5 and 5 (e.g. Speagle et al. 2014). Most studies
assume a linear relation between $log(s\SFR)$ and $log(\Ms)$ and
characterize the slope and the scatter of this relation. Depending on
the survey characteristics and on the $\SFR$ tracer, the value of the
slope varies significantly in the literature. Noeske et al. (2007a)
find a slope of -0.33$^{\pm 0.08}$, while several other studies obtain
a value close to -0.1 (Elbaz et al. 2007, Daddi et al. 2007, Pannella
et al. 2009, Peng et al. 2010). In the compilation of Speagle et
al. (2014), the slope varies between -0.65 and -0.05 and depends on
the $\SFR$ tracer. Some studies show that the slope could depend on
the stellar mass and they even show a probable break in the
$\Ms$-$s\SFR$ relation (e.g. Noeske et al. 2007a, Bauer et al. 2013,
Lee et al. 2015, Whitaker et al. 2014). The parametrization assuming a
linear relation between $log(s\SFR)$ and $log(\Ms)$ is likely not
valid over the full mass range. Finally, the scatter of the
$\Ms$-$s\SFR$ relation is also debated in the literature. The scatter
is ranging from 0.15 dex (Salmi et al. 2012) to 0.5 dex (Salim et
al. 2007) and does not depend on the mass (e.g. Speagle et al. 2014,
Lee et al. 2015). While studied in great detail, no consensus has been
reached on the evolution, the scatter and the slope of the
$\Ms$-$s\SFR$ relation.

Most of the analyses of the $\Ms$-$s\SFR$ relation are based on
scatter diagrams (i.e. displays of the location of the individual
sources in the $\Ms$-$s\SFR$ plane). However, this method does not
provide any quantitative information on how galaxies are distributed
around the median $s\SFR$, and does not account for galaxies that
could be under-sampled or missed by selection effects. In order to
overcome this limitation, one should split the $\Ms$-$s\SFR$ plane in
several mass bins and characterize the $s\SFR$ distribution in each
bin correcting for selection effects. Then, accurate and robust
information can be extracted from the analysis of the $s\SFR$
distribution.

The $s\SFR$ distribution has already been investigated in a few
studies. Guo et al. (2013) produced the $s\SFR$ distributions per
stellar mass bin in a sample similar to ours. Their study is limited
to $0.6<z<0.8$ while we want to explore a large redshift range in
order to analyse the $s\SFR$ evolution. The work of Rodighiero et
al. (2011) is also limited to one redshift slice at $z\sim
2$. Moreover, Rodighiero et al. (2011) need to rely on the UV light to
trace the $s\SFR$ for the bulk of the star-forming
population. Unfortunately, converting the UV light into $\SFR$
introduces uncertainties since it requires an estimate of the UV light
absorbed by dust (e.g. Heinis et al. 2013, Rodighiero et al. 2014). By
fitting a log-normal function over the $s\SFR$ distribution
established by Rodighiero et al. (2011), Sargent et al. (2012) find
$\sigma=0.188^{+0.003}_{-0.003}$. Moreover, Sargent et al. (2012)
include in their fit a population of starbursts, i.e., galaxies having
a $s\SFR$ higher than expected from the main sequence position. They
estimate that 4\% of the galaxies could be considered starbursts at
$z\sim 2$. Finally, Kajisawa et al. (2010) measured the $s\SFR$
distribution per mass bin at $0.5<z<3$ but with a sample limited in
size.

Here, we estimate the $s\SFR$ functions, i.e., the number density of
the galaxy per comoving volume and per $s\SFR$ bin. We measure the
$s\SFR$ functions in four stellar mass bins from $z=0.2$ to
$z=1.4$. In order to overcome the limitations of previous studies, we
follow the following principles. First, our results rely on one robust
$\SFR$ tracer, the 24$\mu m$ IR data obtained with the Multiband
Imaging Photometer (MIPS) camera on board the Spitzer satellite. By
limiting the analysis at $z<1.4$, the galaxy $L_{IR}$ can be derived
with an accuracy better than 0.15 dex using the MIPS 24$\mu m$ data
(Elbaz et al. 2010). The advantage of using a 24$\mu m$ selected
sample is that we reach a lower $\SFR$ limit in comparison to a sample
selected in one Herschel band. Since we apply one single cut in flux,
we can easily correct for selection effect. Second, we limit our
analysis to galaxy samples which are complete in stellar mass. These
criteria allows us to consider only the $\SFR$ limit without having to
consider an additional mass limit. Third, we combine the COSMOS
(Scoville et al. 2007) and GOODS (Giavalisco et al. 2004) surveys. The
large COSMOS area of 1.5 deg$^2$ allows us to get rare and massive
star-forming sources, while the deep GOODS data allow us to study the
shape of the relation at low $s\SFR$ and low mass. Therefore, we have
a broader view of the main sequence and we deal with selection
effects.  Finally, we parametrize the shape of the $s\SFR$ function to
fit the data. We try several options for the parametrization. Based on
these fits, we can derive accurate measurements of the median $s\SFR$,
or of the width of the $s\SFR$ function, which modify some previous
findings on the star-forming main sequence.

The paper is organized as follows. The data are introduced in
\S\ref{data}. Since the ``main sequence'' refers only to star-forming
galaxies, we need to carefully select this population, as described in
\S\ref{starformingSel}.  The method used to estimate the $s\SFR$
functions and the associated uncertainties is explained in
\S\ref{estimate}. We discuss the evolution of the $s\SFR$ functions in
\S\ref{evolGe}. We compare our reference $s\SFR$ functions with the
ones obtained using optical $\SFR$ tracers in \S\ref{opt} and with
predictions of a semi-analytical model in \S\ref{SAM}. Finally, we
discuss our results in \S\ref{discussion} and conclude in
\S\ref{conclusions}.
 
Throughout this paper, we use the standard cosmology
($\Omega_m~=~0.3$, $\Omega_\Lambda~=~0.7$ with
$H_{\rm0}~=~70$~km~s$^{-1}$~Mpc$^{-1}$). Magnitudes are given in the
$AB$ system (Oke 1974). The stellar masses ($\Ms$) are given in units
of solar masses ($\Msol$) for a Chabrier (2003) initial mass function
(IMF). The $s\SFR$ is given in $Gyr^{-1}$.

\begin{figure}[htb!]
\includegraphics[width=8.5cm]{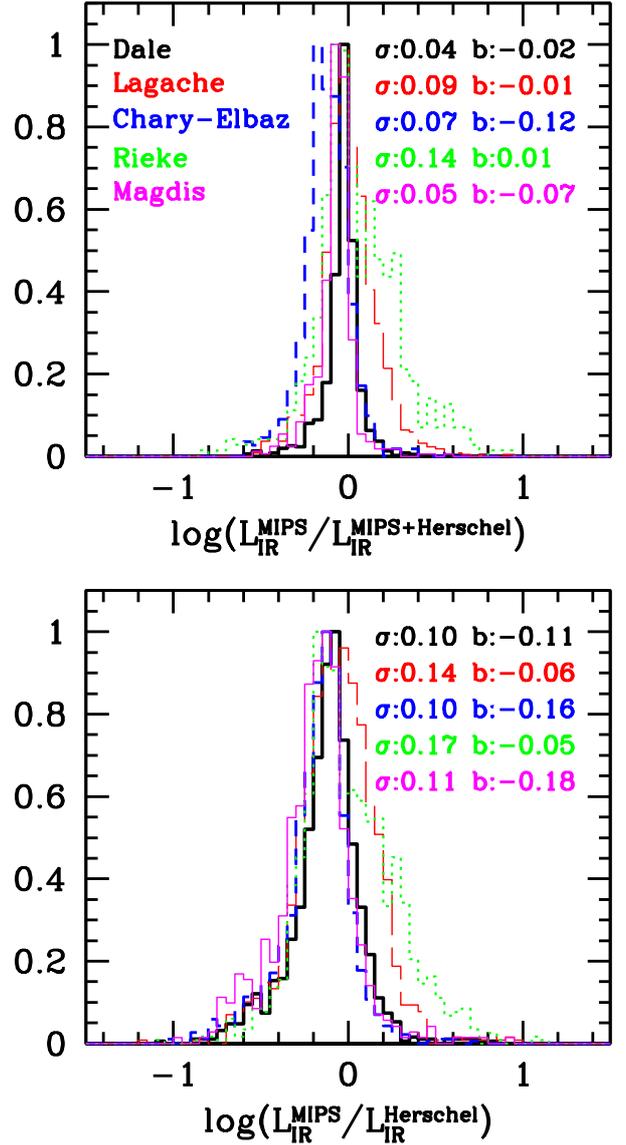} 
\caption{Comparison between the total infrared luminosities derived
  using only the 24$\mu m$ data ($L_{\rm IR}^{\rm MIPS}$), using the
  combination of 24$\mu m$ and Herschel data ($L_{\rm IR}^{\rm
    MIPS+Herschel}$), and using only Herschel data ($L_{\rm IR}^{\rm
    Herschel}$). We indicate in each panel the dispersion ($\sigma$)
  between both measurements and the median ($b$) of the
  distribution. We provide a comparison for several sets of templates
  (Dale \& Helou 2002, Chary \& Elbaz 2001, Rieke et al. 2009, Lagache
  et al. 2004 and B{\'e}thermin et al. 2012 following Magdis et
  al. 2012).
\label{histLIR}}
\end{figure}

\begin{figure*}[htb!]
\begin{center}
\includegraphics[width=14cm]{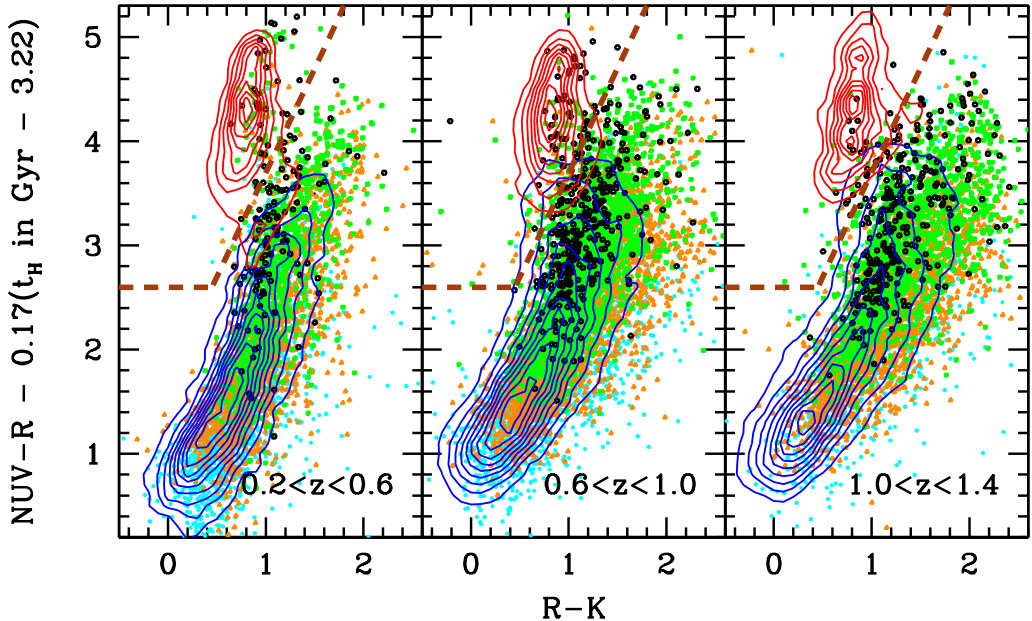} 
\end{center}
\caption{$NUV-R$ versus $R-K$ rest-frame colors in the COSMOS field at
  $0.2<z<1.4$ and $log(\Ms/\Msol)>9.5$. An additional term depending
  on the redshift is added to the $NUV-R$ color in order to keep the
  same criterion to separate quiescent and star-forming galaxies valid
  at all redshifts (brown dashed lines). Cyan crosses, orange
  triangles, green squares and black circles correspond to galaxies
  with masses at $log(\Ms)=9.5-10$, $10-10.5$, $10.5-11$ and
  $11-11.5$, respectively. The red and blue contours indicate the
  distribution of the mass selected galaxies ($log(\Ms/\Msol)>9.5$)
  with $log(s\SFR)_{SED}<-2$ and $log(s\SFR)_{SED}>-2$, respectively.
  \label{NRK}}
\end{figure*}

\section{The galaxy stellar mass and $\SFR$ samples}\label{data}

Our analysis combines the data from the GOODS and the COSMOS surveys
and our measurements are based on MIPS selected samples at
  $F_{24\mu m}>20 \mu Jy$ and $F_{24\mu m}>60 \mu Jy$, respectively.

In the COSMOS field, we use the i$^+$-selected catalogue (limiting
magnitude of 26.2 mag at 5$\sigma$) created by Capak et al. (2007). We
use an updated version of the photometric catalogue including the
UltraVISTA DR1 data release (McCracken et al. 2012) and new SPLASH
IRAC data at 3.6 and 4.5 $\mu m$ (Capak et al., in prep.).  The
photometric redshifts are estimated using 30 bands, as described in
Ilbert et al. (2013). Their accuracy is similar to Ilbert et
al. (2013) in the redshift range considered in this paper
($0.2<z<1.4$). By comparing these photometric redshifts with 10,800
spectroscopic redshifts from the zCOSMOS bright survey (Lilly et
al. 2007), we find a precision of $\sigma_{\Delta z / (1+z)}=0.008$ at
$i^+<22.5$ and $z<1.4$. Using the spectroscopic samples from Comparat
et al. (2015), Capak et al. (2015, in prep.) and the VIMOS Ultra-Deep
Survey (Le F\`evre et al. 2014), we find $\sigma_{\Delta z /
  (1+z)}=0.03$ at $i^+<24$.

The stellar masses are estimated using ``Le Phare'' (Arnouts et
al. 2002, Ilbert et al. 2006). We define the stellar mass as the total
mass in stars at the considered age (without the mass returned to the
interstellar medium by evolved stars). We derive the galaxy stellar
masses using a library of synthetic spectra generated using the
stellar population synthesis (SPS) model of Bruzual and Charlot
(2003). In addition to the library used in Ilbert et al. (2010)
assuming exponentially declining SFH, we add two other star formation
histories based on delayed SFH ($\tau^{-2} t e^{-t/\tau}$) having a
maximum $\SFR$ peak after 1 and 3 Gyr. For all these templates, two
metallicities (solar and half-solar) are considered. Emission lines
are added following Ilbert et al. (2009). We include two attenuation
curves: the starburst curve of Calzetti et al. (2000) and a curve with
a slope $\lambda^{−0.9}$ (Appendix A of Arnouts et al. 2013). E(B-V)
is allowed to take values as high as 0.7. We assign the mass using the
median of the marginalized probability distribution function (PDF). As
shown in Mitchell et al. (2013), this procedure allows us to reduce
some discontinuities in the mass estimate. The 1$\sigma$ uncertainties
derived from the PDF increase from 0.035 dex at $0.2<z<0.4$ to 0.055
dex at $1.2<z<1.4$ for the MIPS selected sample considered in this
paper. We also match our own mass estimates with the two independent
measurements of the masses published in Brammer et al. (2011) and
Muzzin et al. (2013). The three mass catalogues are established for
the same sources, but use a different photometry, different photo-z
codes and different assumptions to construct the SED templates. Based
on this comparison\footnote{The dispersion increases from 0.06 dex at
  $0.2<z<0.4$ to 0.1 dex at $1.2<z<1.4$ between Muzzin et al. (2013)
  and our own stellar masses, with a systematic shift of 0.05
  dex. When considering the Brammer et al. (2011) catalogue, we find a
  similar dispersion and no systematic shift. Since the dispersion
  between two catalogues is the combination of both stellar mass
  uncertainties, the dispersion needs to be divided by $\sqrt 2$ to
  get the real uncertainties.}, we conclude that the mass
uncertainties increase from 0.05 dex at $0.2<z<0.4$ to 0.07 dex at
$1.2<z<1.4$. Systematic uncertainties on the stellar masses (e.g. due
to the IMF or SPS choices) are not included here.

The main part of our analysis is based on a MIPS 24 $\mu m$ selected
catalogue.  The deep MIPS S-COSMOS data were taken during Spitzer
Cycle 3 and cover the full COSMOS 2-deg$^2$ (Sanders et al. 2007). The
$24\mu m$ sources are detected with SExtractor (Bertin \& Arnouts
1996) and their fluxes measured with a PSF fitting technique (Le
Floc'h et al. 2009). Le Floc'h et al. (2009) identified the
optical/NIR counterparts of the $24\mu m$ detection. When possible, we
also use the Herschel data observed as part of the PEP survey at 100
and 160$\mu m$ (Lutz et al.  2011) and Hermes survey at 250, 350 and
500 $\mu m$ (Oliver et al. 2012). The Herschel fluxes are extracted
using the $24\mu m$ catalogue as prior which makes the
cross-identification with the optical sample straightforward.

We estimate the $\SFR$ of our sample following exactly the same method
as Arnouts et al. (2013). The total $\SFR$ is obtained by summing the
contribution of the IR and UV light using Eq.1 of Arnouts et
al. (2013), i.e. $SFR [\Msol yr^{−1}]=8.6 \times 10^{-11} (L_{IR} +
2.3 L_{NUV})$. The infrared luminosities $L_{\rm IR}^{\rm MIPS}$ are
extrapolated from the 24 $\mu m$ fluxes using the Dale \& Helou (2002)
library following Le Floc'h et al. (2009). With this method, a given
IR luminosity is associated to one template. In order to quantify the
uncertainties generated by this extrapolation, we derive $L_{\rm
  IR}^{\rm MIPS+Herschel}$ using a minimum of three bands (the 24 $\mu
m$, one band from PACS and one from SPIRE) and we allow any template
to be fitted. As shown in the top panel of Fig.\ref{histLIR}, we find
no systematic offset between $L_{\rm IR}^{\rm MIPS}$ and $L_{\rm
  IR}^{\rm MIPS+Herschel}$. The dispersion between both measurements
increases from 0.03 dex at $z<0.6$ to 0.12 at $z>1$. Therefore, our
extrapolation from the 24 $\mu m$ flux assuming one SED for a given
$L_{\rm IR}$ does not introduce significant uncertainties or
biases. We also compare $L_{\rm IR}^{MIPS}$ and $L_{\rm IR}^{\rm
  Herschel}$ ($L_{\rm IR}^{\rm Herschel}$ is computed without using
the 24 $\mu m$ data) in order to use two independent estimates of the
$L_{\rm IR}$. Based on the scatter of $L_{\rm IR}^{MIPS} - L_{\rm
  IR}^{\rm Herschel}$, we expect an uncertainty on the $L_{\rm IR}$ of
0.06 dex, 0.09 dex, and 0.13 dex at $0.2<z<0.6$, $0.6<z<1.0$, and
$1<z<1.4$, respectively. We observe a systematic offset of 0.1 dex
between $L_{\rm IR}^{\rm Herschel}$ and $L_{\rm IR}^{\rm MIPS}$,
showing that one of the two estimates could be biased. As shown in
Fig.\ref{histLIR}, this offset is present for several sets of
templates available in the literature. Such an offset could be
partially explained by the combined uncertainties in the absolute
calibration of MIPS and/or Herschel data\footnote{A preliminary
  reduction of the MIPS data using the S18.0 SSC pipeline rather than
  the S12 pipeline show that a possible factor of 1.1 should be
  applied to our measured flux at 24$\mu m$ (Aussel et al., private
  communication). Moreover, we observe differences reaching 30\%
  between the 24$\mu m$ fluxes in published catalogues from the
  literature in GOODS and COSMOS (Wuyts et al. 2008, Le Floc'h et
  al. 2009, Muzzin et al. 2013, Magnelli et al. 2013). It shows that
  an uncertainty in the 24$\mu m$ total fluxes is plausible. Moreover,
  uncertainties in the absolute calibration of MIPS and Herschel are
  combined in this comparison.}. Still, a systematic shift independent
of the redshift does not affect our conclusions.  Hereafter, we adopt
the Dale \& Helou (2002) templates and we use $L_{\rm IR}^{\rm
  MIPS+Herschel}$ to get the total infrared luminosity.

Since AGN could contaminate the 24$\mu m$ emission and bias the
stellar mass estimate, we remove the bright X-ray sources detected in
XMM (Brusa et al. 2007). We keep the sources identified as IRAC
power-laws (Donley et al. 2012), but we also perform the full analysis
removing the IRAC power-laws without a noticeable change in our
results.\\
 
In the GOODS field, we use the FIREWORKS data published by Wuyts et
al. (2008). This catalogue reaches $K<24.3$ at 5$\sigma$ over 138
arcmin$^2$. We compute the photometric redshifts using Le Phare and
the same method as the COSMOS field. We obtain photometric redshifts
comparable to the ones of Wuyts et al. (2008) with a precision at
$\sigma_{\Delta z / (1+z)}=0.03$ at $i^+<24$. The comparison is based
on several spec-z samples compiled by Wuyts et al. (2008), including
VVDS data from Le F\`evre et al. (2004) and K20 data from Mignoli et
al. (2005). We apply exactly the same method as in COSMOS to derive
the stellar masses.  This catalogue also includes MIPS data. Following
Wuyts et al. (2008), we apply a selection at $F_{24\mu m}>20 \mu Jy$
in this catalogue.  We also add the GOODS-Herschel data at 100 and
160$\mu m$ (Elbaz et al, 2011). The $\SFR$ is estimated following
exactly the same method as for the COSMOS field.

\begin{figure}[htb!]
\centering \includegraphics[width=9.cm]{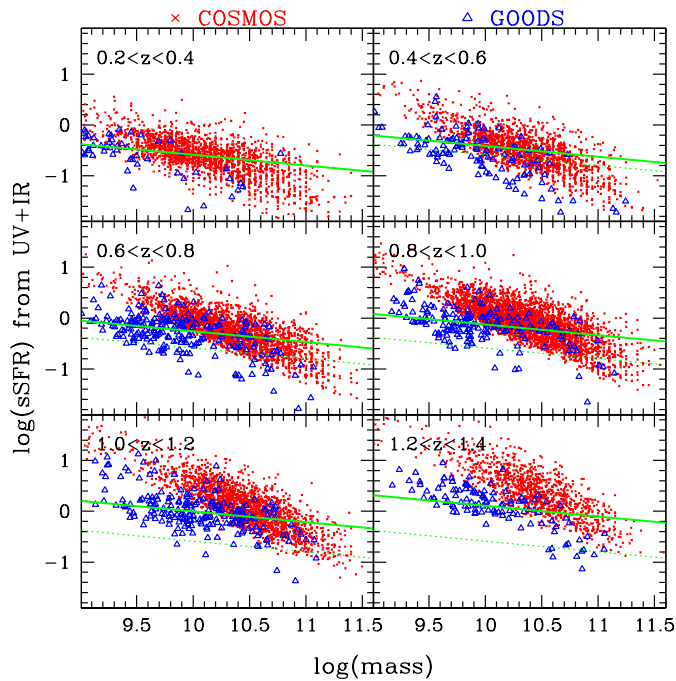}
\caption{$s\SFR$ as a function of the stellar mass in the GOODS (blue
  triangles) and COSMOS (red crosses) fields with the $\SFR$ measured
  from the UV and IR data. The green dashed lines are obtained using
  the parametrization obtained by Sargent et al. (2012). The green
  dashed line corresponds to the relation at $0.2<z<0.4$.}
           \label{masssfr} 
\end{figure}

\section{Selecting the star-forming galaxies}\label{starformingSel}

In order to study the evolution of the main sequence, we need to
identify star-forming and quiescent galaxies.  The presence of a
bimodal distribution in color (e.g., Bell et al. 2004, Faber et
al. 2007, Franzetti et al. 2007, Smol{\v c}i{\'c} et al. 2008, Fritz
et al. 2014) or in the $\Ms$-$\SFR$ plane (e.g., Peng et al. 2010)
shows that galaxies are transitioning rapidly from a star-forming main
sequence to a red clump. Therefore, a quantitative criterion can be
established to select the star-forming galaxies.

Williams et al. (2009) show that the combination of two rest-frame
colors ($M_U-M_V$, $M_V-M_J$) is sufficient to separate quiescent and
star-forming galaxies without mixing galaxies that are red because of
dust extinction and the ones with a quenched star formation. We use a
modified version of this selection criterion by combining the two
rest-frame colors $M_{NUV}-M_R$ and $M_R-M_K$ following Arnouts et
al. (2013). The absolute magnitudes are derived using the method
described in Appendix B of Ilbert et al. (2005): in order to minimize
the uncertainty induced by the k-correction term, the rest-frame
luminosity at a given wavelength $\lambda$ is derived from the
apparent magnitude observed at $\lambda (1 + z)$. Figure \ref{NRK}
shows the galaxy distribution in the NUV-R-K plane within the COSMOS
field. The red clump is clearly isolated from the star-forming
sequence by a lower density region in which galaxies transit rapidly.
We establish a limit to separate the quiescent and the star-forming
galaxies within this lower density region in the NUV-R-K plane. This
limit changes with cosmic time because of the evolution of the stellar
populations. In order to apply a single criterion at all redshifts to
select the star-forming galaxies, we add a time dependent correction
$C$ to our selection criterion, with $C=-0.17[t_H(z)-t_H(z=2)]$ if
$z<2$ and $t_H$ the age of the Universe at a given redshift in
Gyr. The galaxies with $(M_{NUV}-M_R)+C<2.6$ and
$(M_{NUV}-M_R)+C<2(M_R-M_K)+1.7$ are considered to be
star-forming. The considered limit is indicated with the brown dashed
lines in Fig.\ref{NRK}. We note that the time correction $C$ is
established empirically to produce the cleanest separation between the
red and blue regions\footnote{Having a theoretical justification of
  the correction is not possible since we do not know the mix of SFH
  and the ages of the galaxies around the transitioning area. Still,
  by using a BC03 template with an exponentially declining SFH and
  $\tau=3Gyr$ ($\tau=2Gyr$), we would get a color correction of
  approximately -0.33 (-0.2) in $(M_{NUV}-M_R)$ and -0.05 (-0.06) in
  $(M_R-M_K)$. Our applied empirical correction $C$ falls in this
  range.}. The same criterion is applied to the GOODS sample,
providing an equally good separation between the star-forming sequence
and the red clump.

We find that 2\% of the MIPS sources fall within the quiescent region
at $0.2<z<1.4$ (above the brown dashed lines). This 24$\mu m$ flux
could be explained by the contribution of an AGN. Moreover, some
post-starburst galaxies with a quenched star formation could still be
seen in IR (e.g. Hayward et al. 2014). Therefore, we do not include
the sources falling in the quiescent locus in our analysis. However,
we have checked that including this population would not affect our
conclusions.

We emphasize the importance of using a two-color criterion to study
the $\Ms$-$\SFR$ relation for the most massive galaxies. We would
remove a significant fraction of massive dust-extinguished
star-forming galaxies from the main sequence by using a single-color
criterion. For instance, we would loose 20\% of the galaxies more
massive than $\Ms>10^{10.5}\Msol$ with a selection
$(M_{NUV}-M_R)>3.5$.

Finally, our color-color selection corresponds approximately to a cut
in $log(s\SFR_{SED})$ at $-2$ (e.g. Ilbert et al. 2010) with
$s\SFR_{SED}$ estimated using the template fitting procedure (blue and
red contours in Fig.\ref{NRK}). We check with our dataset that 3-4\%
of all our star-forming sources at $\Ms>10^{9.5}\Msol$ (not MIPS
selected) have $log(s\SFR_{SED})<-2$ while 3-4\% of the galaxies that
we do not classified as star-forming have $s\SFR_{SED}>-2$. The
majority of these sources are located $\pm 1 dex$ around
$log(s\SFR)=-2$. Therefore, our classification in colors is very
similar to a classification in $s\SFR_{SED}$.

\begin{figure*}[htb!]
\centering \includegraphics[width=19.cm]{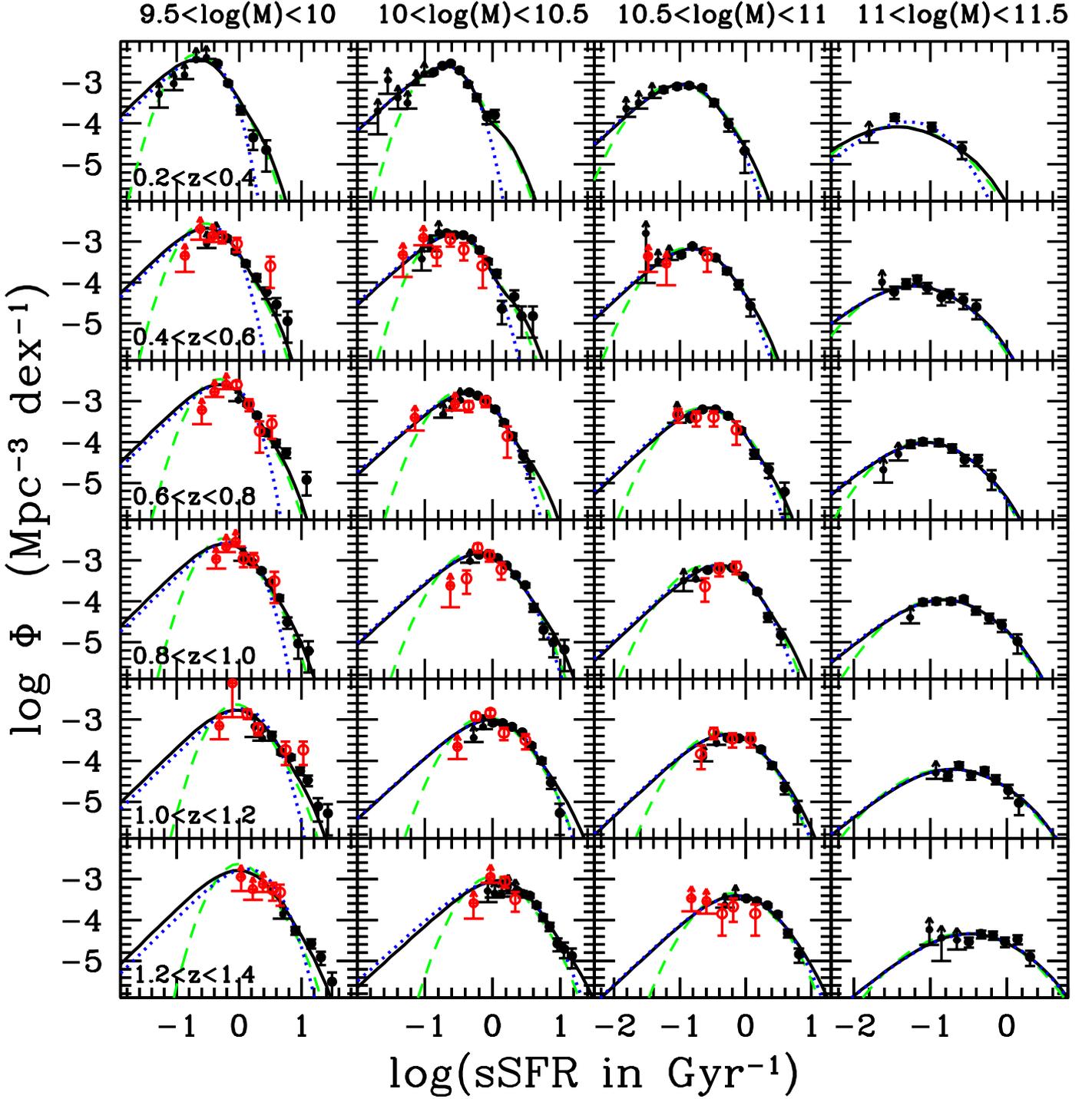} 
\caption{$s\SFR$ functions per redshift bin from $0.2<z<0.4$ to
  $1.2<z<1.4$ (from the top to the bottom rows) and per stellar mass
  bin from $9.5<log(\Ms)<10$ to $11<log(\Ms)<11.5$ (from the left to
  the right columns). The non-parametric data have been obtained using
  the 1/V$_{\rm max}$ estimator. The black filled and red open circles
  correspond to the COSMOS and GOODS fields, respectively. The arrows
  correspond to the lower limits obtained with the 1/V$_{\rm
    max}$. The black solid lines and green dashed lines correspond to
  the best-fit functions assuming a double-exponential and a
  log-normal profile, respectively. Both include a starburst component
  (see details in \S\ref{methodsSFR}). The blue dotted lines
  correspond to the double-exponential fit without considering the
  starburst component.
  \label{allLF}}
\end{figure*}

\section{Measurement and fit of the $s\SFR$ functions}\label{estimate}

In this section, we describe the method used to derive the $s\SFR$
functions per stellar mass bin. We discuss the possible selection
effects {in the mass-$s\SFR$ plan shown in Fig.\ref{masssfr}} and
correct for them when necessary (mainly the 24$\mu m$ flux limit since
we are complete in stellar mass). We assume two possible profiles for
the $s\SFR$ functions (a log-normal function and a double exponential
function). Figure \ref{allLF} shows the $s\SFR$ functions, and the
best-fit parameters are given in Tables \ref{paraDoubleExpo} and
\ref{paraGauss}.

\subsection{The mass-$s\SFR$ scatter diagram}\label{scatter}

Figure \ref{masssfr} shows the distribution of the $s\SFR$ as a function
of the stellar mass for star-forming galaxies in the COSMOS field (red
crosses) and in the GOODS field (blue triangles). Since GOODS covers a
small volume with a deep NIR coverage, this sample includes
preferentially low-mass galaxies at $z<1$, while COSMOS which covers
an area $\times 30$ larger includes rare and massive sources. This
difference explains why the GOODS and the COSMOS samples cover a
different mass range in Fig.\ref{masssfr}. Still, the $s\SFR$ values
of the COSMOS survey are larger than the values found in the GOODS
field for masses $\Ms<10^{10}\Msol$. This effect is explained by the
$\times 3$ difference in sensitivity between the two MIPS
surveys. While the COSMOS survey includes mostly starbursting sources
at low masses, the GOODS survey is able to reach the bulk of the
star-forming population. We will discuss in more detail this selection
effect in \S\ref{SAM} using a semi-analytical model.

The green solid line corresponds to the relation
\begin{equation}
log(s\SFR)= -7.81 - 0.21 \times log(M^*) +2.8\times log(1+z) 
\end{equation}
established using the mass dependency of the $s\SFR$ of $-0.21$
provided by Rodighiero et al. (2011), as well as the normalization of
the main sequence at $z\sim 2$ from the same analysis. We assume an
evolution in $(1+z)^{2.8}$ from Sargent et al. (2012). The position of
our GOODS data agrees with this relation. However, such
parametrization is not suitable for the COSMOS field. It demonstrates
the need for a statistical study taking into account 
selection effects.

\subsection{Methodology to estimate the non-parametric $s\SFR$ functions}\label{methodsSFR}

In order to fully characterize the evolution and the shape of the main
sequence, we measure the $s\SFR$ function, i.e., the number density in
a comoving volume (in Mpc$^{-3}$) and per logarithmic bin of $s\SFR$
(in dex$^{-1}$). We derive the $s\SFR$ function per stellar mass and
redshift bin. We divide the star-forming sample into 6 redshift bins
with $\Delta z=0.2$ and four stellar mass bins $log(\Ms)=9.5-10$ dex,
$10-10.5$, $10.5-11$, $11-11.5$.

We note that the $s\SFR$ functions are measured per stellar mass bin
of 0.5 dex. Therefore, one has to multiply their normalization by 2 in
order to express the density per logarithmic bin of $s\SFR$ and per
logarithmic bin of $\Ms$ simultaneously, i.e., in Mpc$^{-3}$dex$^{-2}$
(a bivariable galaxy mass and $s\SFR$ function).

In order to take into account the flux limit at 24$\mu m$ ($F_{24\mu
  m}>20\mu Jy$ in GOODS and $F_{24\mu m}>60\mu Jy$ in COSMOS), we
adopt standard estimators as the 1/V$_{\rm max}$ (Schmidt 1968), the
SWML (Efstathiou 1988) and the C$^+$ (Lynden-Bell 1971). These
estimators are included in the tool ALF used to compute the $s\SFR$
function, as described in Appendix B of Ilbert et al. (2005).

Because of the depth of the COSMOS optical and NIR images, we do not
need to consider any incompleteness in stellar mass.  Indeed, only 4\%
and 0.5\% of the galaxies are fainter than $i>25.5$ and $m(3.6)>24$
(this magnitude limits are 0.5-1 mag brighter than the magnitude limit
of our survey) in the most incomplete bin $\Ms<10^{10}\Msol$ and
$z>1.2$. Only 2\% of the star-forming galaxies would require a
1/V$_{\rm max}$ correction in this bin for our considered limits in
NIR. Therefore, the samples considered in this analysis are complete
in mass. Since the GOODS data are deeper than the COSMOS data in
optical, the GOODS sample is also complete in stellar mass at
$\Ms>10^{9.5}\Msol$ and $z<1.4$.

We define $s\SFR$ limits, denoted $s\SFR_{complete}$, above which we
can safely correct for selection effects. As shown in Ilbert et
al. (2004), if a galaxy population is not observable anymore below a
given $s\SFR$, denoted $s\SFR_{complete}$, the standard estimators
cannot correct for this missing population. Moreover, the various
estimators are biased differently below $s\SFR_{complete}$. We adopt
the following definition: $s\SFR_{complete}$ is the $s\SFR$ for which
90\% of the galaxies have their $s\SFR_{limit}<s\SFR_{complete}$, with
$s\SFR_{limit}$ being the lowest $s\SFR$ observable for each galaxy
given the 24$\mu m$ flux limit. Following this procedure, not more
than 10\% of the galaxies could be missed at
$s\SFR>s\SFR_{complete}$. We also restrict our analysis to the $s\SFR$
range where the 3 non-parametric estimators produce consistent
results. We will use the 1/V$_{\rm max}$ estimator in this paper, but
the results would be the same at $s\SFR>s\SFR_{complete}$ using the
other estimators. One advantage of the 1/V$_{\rm max}$ estimator is
that it produces a lower limit in density at $s\SFR<s\SFR_{complete}$
(Ilbert et al. 2004). Therefore, we conserve this important
information and use the 1/V$_{\rm max}$ estimator as lower limits when
possible (the lower limits are shown with arrows in Fig. \ref{allLF}).

Star-forming sources with an extreme dust attenuation could be missing
in the parent optical photometric catalogue. Indeed, Le Floc'h et
al. (2009) do not find any optical counterpart for 10$\%$ of the MIPS
sources. Given our survey limits, we establish that these missing
sources are redder than $m_R-m_{24}>7.6$ (or fake MIPS
detections). Dey et al. (2008) and Riguccini et al. (2011) show that
$>85\%$ galaxies selected with such color criteria are located at
$z>1.4$, i.e., above the maximum redshift considered in our study.

Finally, the GOODS field (138 arcmin$^2$) covers a smaller area than the
COSMOS field (1.5 deg$^2$), the uncertainties due to the cosmic
variance reach $\sigma=0.21-0.37$ (for $log(\Ms)=9.75-11.25$) at $z\sim
0.9$ in the GOODS field and $\sigma=0.09-0.15$ in the COSMOS field
(Moster et al. 2011). Therefore, we use COSMOS as the anchor for the
normalization of the $s\SFR$ functions. We adjust the normalization of
the $s\SFR$ functions in the GOODS field by applying the following
factors to the normalization: 1.243, 0.7653, 0.7605, 1.305, 0.9264 and
1.186 in the redshift bins 0.2-0.4, 0.4-0.6, 0.6-0.8, 0.8-1, 1-1.2 and
1.2-1.4, respectively. These factors are derived from the ratio
between the redshift distributions of the two fields for a same
magnitude selection limit.

\begin{figure*}[htb!]
\begin{tabular}{c c}
\includegraphics[width=8.5cm]{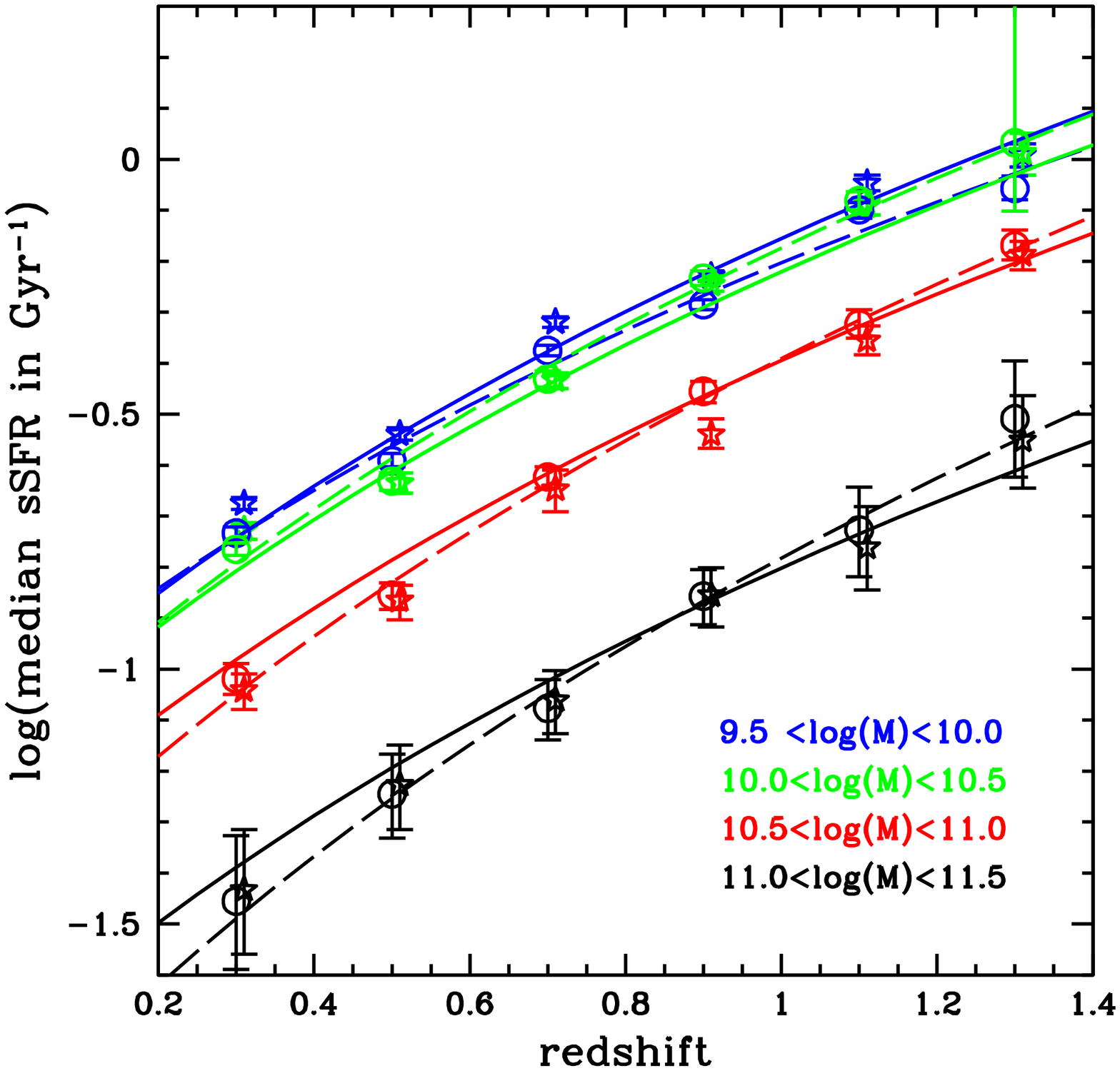} &
\includegraphics[width=8.5cm]{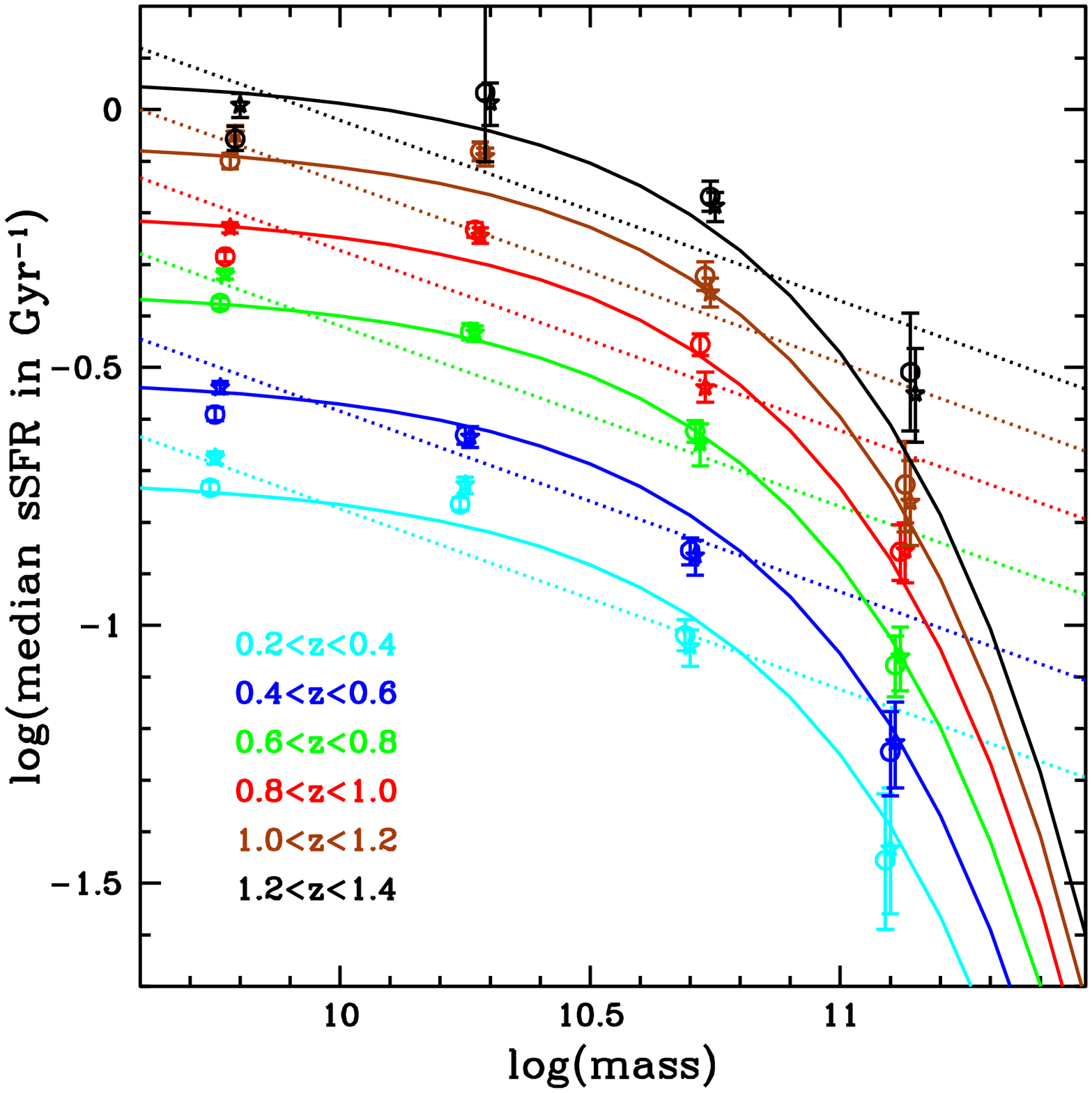} 
\end{tabular}
\caption{Evolution of the median $s\SFR$ as a function of redshift
  (left panel) and stellar mass (right panel). Open stars and open
  circles correspond to the values measured assuming a log-normal and
  a double-exponential profiles, respectively. The solid lines
  correspond to the fit using Eq.\ref{ssfrEvol} and $b$ independent of
  $\Ms$. {\it Left:} each color corresponds to a stellar mass bin
  (blue: $9.5-10$ dex, green: $10-10.5$, red: $10.5-11$ and black
  $11-11.5$). The dashed lines are obtained with $b$ varying in each
  mass bin. {\it Right:} each color corresponds to a redshift bin from
  $0.2-0.4$ (cyan) to $1.2-1.4$ (black). The dotted lines are obtained
  using $log(s\SFR) \propto log(\Ms)$ (i.e., the standard definition in
  the literature). The solid line corresponds to $log(s\SFR) \propto
  -0.17\Ms$
  \label{sSFRevol}}
\end{figure*}

\subsection{Parametric fit of the  $s\SFR$  functions}\label{fit}

We fit simultaneously the 1/V$_{\rm max}$ data of the COSMOS and GOODS
fields. We consider two possible profiles: a log-normal function and a
double exponential function. The log-normal function is parametrized
as:
\begin{equation}
\phi(s\SFR)=\frac{\Phi^*}{\sigma \sqrt{2\pi}}exp(-\frac{log_{10}^2(s\SFR / s\SFR^*)}{2\sigma^2})
\end{equation}
with $\Phi^*$ the normalization factor, $s\SFR^*$ the characteristic
$s\SFR$, and $\sigma$ the standard deviation. We also consider a
double-exponential profile (e.g., Saunders et al. 1990, Le Floc'h et
al. 2005),
\begin{equation}
\phi(s\SFR)=\Phi^* \left( \frac{s\SFR}{s\SFR^*} \right) ^{1-\alpha} exp( -\frac{log_{10}^2(1+ \frac{s\SFR}{s\SFR^*})}{2\sigma^2})
\end{equation}
with $\alpha$ the faint-end slope. While the double-exponential
profile is not commonly used to describe the $s\SFR$ distribution, it
allows for a significant density of star-forming galaxies with a low
$s\SFR$.

In order to take into account the uncertainties on the $s\SFR$, we
convolve these profiles with a Gaussian function having a standard
deviation $\sigma=0.06$, 0.07, 0.08, 0.11, 0.14 and 0.17 dex at
$z=0.2-0.4$, $z=0.4-0.6$, $z=0.6-0.8$, $z=0.8-1.0$, $z=1.0-1.2$, and
$z=1.2-1.4$, respectively. These values are obtained by summing in
quadrature the statistical uncertainties expected for the $\SFR$ and
the stellar masses, as estimated in \S\ref{data} (systematic
uncertainties are not included). We note that $\phi_{c}$ is the
convolved profile.

We add a starburst component to the $s\SFR$ function to produce a
better fit of the 1/V$_{\rm max}$ data at high $s\SFR$.  Since the
contribution of the starbursts cannot be constrained in each
individual bin of redshift and stellar mass, we set their contribution
following Sargent et al. (2012). We assume that the starbursts are
distributed with a log-normal distribution having $\sigma=0.25$ and
centered on the median $s\SFR$ shifted by $+0.6$ dex. We set the
starburst contribution to be 3\% of the total density of star-forming
galaxies.

We fit the 1/V$_{\rm max}$ data measured in the stellar mass bin
[$\Ms^{min};\Ms^{max}$] by minimizing the $\chi^2$ value defined as:
\begin{equation}\label{chi2}
\begin{split}
\chi^2  = & \sum_{i=1,N}^{e_{\Phi_i^{V_{max}}}>0} \left(
\frac{\phi_{c}(s\SFR_i) - \Phi_i^{V_{max}}}{e_{\Phi_i^{V_{max}}}}
\right)^2 \\ 
             & + \sum_{i=1,N}^{\substack{e_{\Phi_i^{V_{max}}}<0 \; \& \; \\ 
  \phi_{c}(s\SFR_i)<\Phi_i^{V_{max}}}} \left(\frac{\phi_{c}(s\SFR_i) -
  \Phi_i^{V_{max}}}{e_{\Phi_i^{V_{max}}}} \right)^2\\ 
            & + \left(
\frac{\int^\infty_{0}\phi_{c}(s\SFR)d(s\SFR) - \MF \times \Delta
  M}{e_\MF}\right)^2
\end{split}
\end{equation}
with $\phi_c$ being the function that we fit, $\Phi_i^{V_{max}}$ the
density at $s\SFR_i$ estimated with the 1/V$_{\rm max}$ estimator, and
$e_{\Phi_i^{V_{max}}}$ its associated Poisson errors; $N$ corresponds
to the number of bins in $s\SFR$. Equation \ref{chi2} contains three
components:
\begin{itemize}
\item the first term of the equation corresponds
to the standard $\chi^2$ minimization method;
\item the second term accounts for the lower limits obtained with the
  1/V$_{\rm max}$ estimator below the completeness limit. The negative
  error $e_{\Phi_i}^{V_{max}}$ indicates that the density is a
  lower limit;
\item the third term includes an additional constraint using the
  galaxy stellar mass function of the star-forming galaxies, denoted
  $\MF$. Indeed, the $s\SFR$ function integrated over the full $s\SFR$
  range should match the $\MF$ integrated over the considered mass bin
  $[log(\Ms^{min}),log(\Ms^{max})]$ with $\Delta M =
  log(\Ms^{max})-log(\Ms^{min})$. Since we slightly modified the
  method used Ilbert et al. (2013) to compute the stellar masses, and
  since we are not using the same parent photometric catalogue, we
  recompute the $\MF$ in each mass/redshift bin with this sample for
  the sake of consistency.
\end{itemize}

The best-fit parameters are given in Tables \ref{paraDoubleExpo} and
\ref{paraGauss} for the double-exponential and the log-normal
functions, respectively. We also add the values of the median and
average $s\SFR$ in these tables. We caution that the median and
average $s\SFR$ are different, even for a log-normal function.

We find that adding the starburst component is necessary in the mass
bin $9.5<log(\Ms)<10$ to reproduce the high $s\SFR$ end, while we
could ignore it above $log(\Ms)>10.5$ (black solid lines and blue
dotted lines in Fig.\ref{allLF}).

Despite the combination of GOODS and COSMOS data, we are not able to
directly constrain the full shape of the $s\SFR$ function. In most of
the redshift and mass bins, the $s\SFR$ function is incomplete below
the peak in $s\SFR$. The lower limits obtained with the 1/V$_{\rm
  max}$ estimator at low $s\SFR$ (the arrows in Fig.\ref{allLF})
indicate the minimum possible densities, which is important in order
to discriminate between a log-normal and a double-exponential
profile. In most of the mass and redshift bins, we do not sample
sufficiently low $s\SFR$ to see any advantage of using either one or
the other parametrization. The fit with a double-exponential function
is more suitable than the log-normal function at $0.2<z<0.6$ and
$10<log(\Ms)<11$ (see Fig.\ref{allLF}). In these bins, the lower
limits favor a double-exponential parametrization. Moreover, the
position of the best-fit $s\SFR$ peak is in better agreement with the
non-parametric data using a double-exponential profile.

Adding the $\MF$ information into Eq.\ref{chi2} brings an important
constraint on the $s\SFR$ distribution, not visible by a simple
examination of the fit. For instance, a higher value of the slope
$\alpha$ of the $s\SFR$ function in the case of a double-exponential
fit will overproduce the density of star-forming galaxies. Even with
this constraint, the uncertainties on $\alpha$ remain large and
$\alpha$ varies between -1 and 0.5 when left free. Therefore, we
arbitrarily set its value at $\alpha=-0.5$ which is suitable in all
the mass/redshift bins.

\section{Evolution of the $s\SFR$ functions}\label{evolGe}

In this section, we analyze the evolution of the median $s\SFR$
derived from the $s\SFR$ functions obtained in \S\ref{fit}. We obtain
a mass-dependent increase in the $s\SFR$ as a function of redshift and
a decrease in $log(s\SFR)$ as $-0.17\Ms$. We also combine all our
$s\SFR$ functions correcting for time evolution, showing that the
width of the $s\SFR$ distribution increases with the stellar mass.

\subsection{Evolution of the median $s\SFR$}\label{evol}

Figure \ref{sSFRevol} shows the evolution of the median $s\SFR$. The
median $s\SFR$ is obtained from the best-fit functions (see
\S\ref{fit}) to avoid selection biases. We observe a clear increase in
the $s\SFR$ as a function of redshift (left panel) and a decrease with
$\Ms$ (right panel).

We adopt the following parametrization of the $s\SFR$ evolution as a
function of redshift and mass,
\begin{equation}\label{ssfrEvol}
log(s\SFR)= a + \beta \times \frac{ \Ms}{ 10^{10.5}\Msol} + b
\times log_{10}(1+z)
\end{equation}
with $a$ the normalization, $\beta$ the dependency on the mass and $b$
the dependency on redshift. The best-fit parameters are given in Table
\ref{parametrisation}.  Assuming that the redshift evolution of the
$s\SFR$ does not depend on the mass, we find $\beta=-0.172\pm 0.007$
and $b=3.14\pm 0.07$. The result is shown with solid lines in
Fig.\ref{sSFRevol}. Then, we relax the assumption that the parameter
$b$ is independent of the stellar mass and we fit each stellar mass
bin independently. We obtain $b=2.88\pm 0.12$, $b=3.31\pm 0.10$,
$b=3.52\pm 0.15$ and $b=3.78\pm 0.60$ in the stellar mass bins
$log(\Ms)=9.5-10$ dex, $10-10.5$, $10.5-11$, and $11-11.5$,
respectively. The result is shown with dashed lines in
Fig.\ref{sSFRevol} (left). It suggests that the evolution is faster
for the massive galaxies, which is in agreement with a downsizing
pattern (Cowie et al. 1996). These values are obtained assuming a
double-exponential profile, but the results are similar if we consider
a log-normal profile.

The parameter $b$ is directly comparable with several values from the
literature using the same parametrization in $\propto (1+z)^b$. In
Karim et al. (2011), $b$ varies between 3.42 and 3.62 at
$10.2<log(\Ms)<11.1$ which is consistent with our results in the same
mass range. Elbaz et al. (2011) find an evolution with $b=2.8$, based
on deep GOODS data. Therefore, a dependency of $b$ on the stellar mass
could explain the differences between the various values found in the
literature.

With our parametrization, $log(s\SFR)$ is proportional to $\Ms$. The
parameter $\beta$ that we obtain cannot be directly compared with
values from the literature. In most of the studies, a linear
dependency with $log(\Ms)$ is assumed (dotted lines in
Fig.\ref{sSFRevol}, right panel) while we assume a linear dependency
with ${\Ms}$ (solid lines). Our parametrization in ${\Ms}$ produces a
more rapid decrease in the $s\SFR$ at $\Ms>10^{10.5}\Msol$ and less
evolution at lower mass than a parametrization in $log(\Ms)$.  As
shown in Fig.\ref{sSFRevol}, a parametrization in $log({\Ms})$ is not
suitable for the considered mass range. It could explain why the slope
values in the literature depends on the considered mass range when
$log(s\SFR) \propto log(\Ms)$ (Lee et al. 2015, Whitaker et
al. 2014). We emphasize that the choice of having $log(s\SFR)$
proportional to $\Ms$ is physically motivated since ``$\tau$ models''
converge toward such parametrization at high masses (e.g. Noeske et
al. 2007b, Bauer et al. 2013).

\begin{figure}[htb!]
\includegraphics[width=8.5cm]{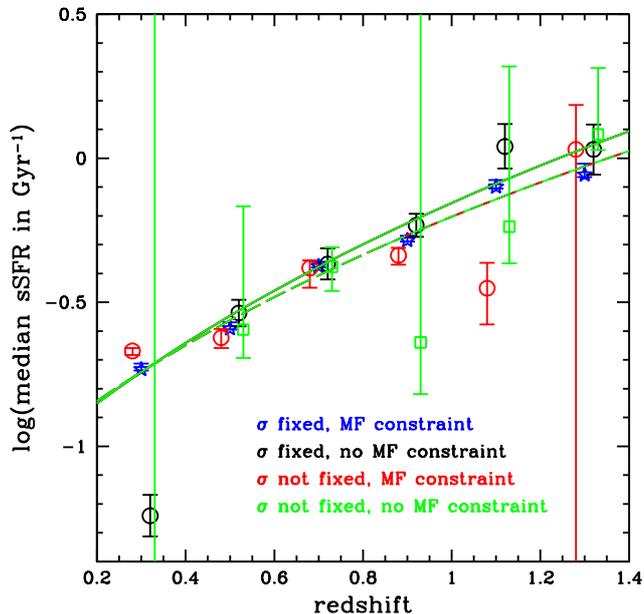} 
\caption{Evolution of the median $s\SFR$ as a function of redshift in
  the stellar mass bin $9.5<log(M)<10$.  The fit is done with the
  double-exponential profile and different options. {\it green:}
  $\sigma$ is free and the $\MF$ constraint is not used; {\it red:}
  $\sigma$ is free and the $\MF$ constraint is used; {\it black:}
  $\sigma$ is fixed and the $\MF$ constraint is not used; {\it blue:}
  $\sigma$ is fixed the $\MF$ constraint is used (default
  configuration).
  \label{sSFR_MF}}
\end{figure}

Finally, we would like to note that including the $\MF$ within the
$\chi^2$ expression (Eq.\ref{chi2}) brings a decisive constraint over
the median $s\SFR$. We illustrate this point in Fig.\ref{sSFR_MF},
showing the evolution of the median sSFR with redshift in the mass bin
$9.5<log(\Ms)<10$. In this mass bin, the non-parametric $s\SFR$
function is primarily composed of lower limits below the $s\SFR$
peak. By using only the information contained in the non-parametric
data to perform the fit, we would not constrain the median $s\SFR$, as
shown by the green error bars. Adding the $\MF$ as an additional
observable within the $\chi^2$ expression breaks the degeneracy
between the best-fit parameters and allows an accurate estimate of the
median $s\SFR$ (red and blue error bars).

\begin{figure}[htb!]
\includegraphics[width=8.5cm]{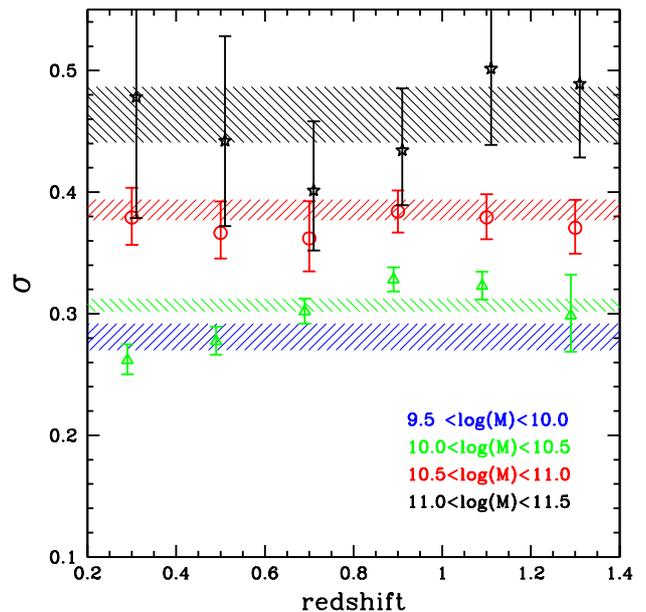} 
\caption{Evolution of the parameter $\sigma$ as a function of
  redshift, obtained by fitting a log-normal function to the 1/V$_{\rm
    max}$ data. Each color corresponds to a stellar mass bin (blue:
  $9.5<log(\Ms)<10$, green: 10-10.5, red: 10.5-11 and black
  11-11.5). The shaded areas correspond to the value measured when all
  the $s\SFR$ functions are combined at $z=0$ as shown in
  Fig.\ref{LFz0}. The individual $\sigma$ points are not measured at
  $9.5<log(\Ms)<10$ since we set the value of $\sigma$ in this mass
  range.
  \label{sigma}}
\end{figure}

\begin{figure*}[htb!]
\centering \includegraphics[width=15cm]{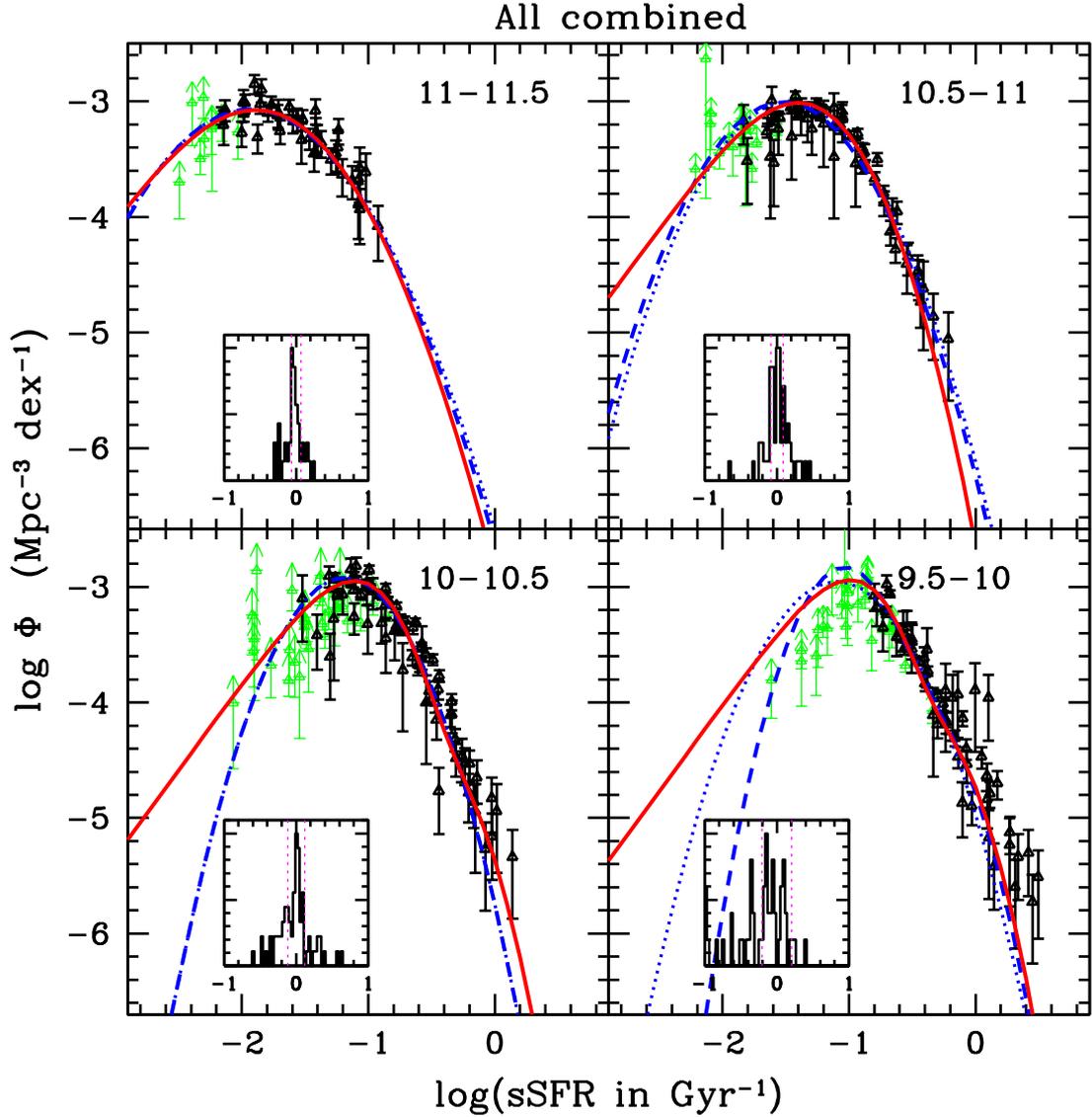} 
\caption{$s\SFR$ functions combined at $z=0$ correcting the 1/V$_{\rm
    max}$ data from the redshift evolution derived in
  \S\ref{fit}. Each panel corresponds to a stellar mass bin. The black
  triangles are obtained with the 1/V$_{\rm max}$ estimator over the
  COSMOS and the GOODS fields (mixed together in this figure). The
  green arrows are lower limits in the 1/V$_{\rm max}$ estimate. The
  red solid lines and blue dashed lines correspond to the best-fit of
  a double-exponential and a log-normal function over the 1/V$_{\rm
    max}$ data. The blue dotted line corresponds to the fit of a
  Gaussian function without including a starburst component. The inset
  in each panel shows the distribution of the differences between the
  best-fit function and the data, with a density dispersion of 0.19
  dex at $log(\Ms)=9.5-10$, 0.11 dex at $log(\Ms)=10-10.5$, 0.09 dex
  at $log(\Ms)=10.5-11$, and 0.07 dex at $log(\Ms)=11-11.5$, as shown
  by the vertical dashed lines.
  \label{LFz0}}
\end{figure*}

\begin{figure*}[htb!]
\includegraphics[width=19.cm]{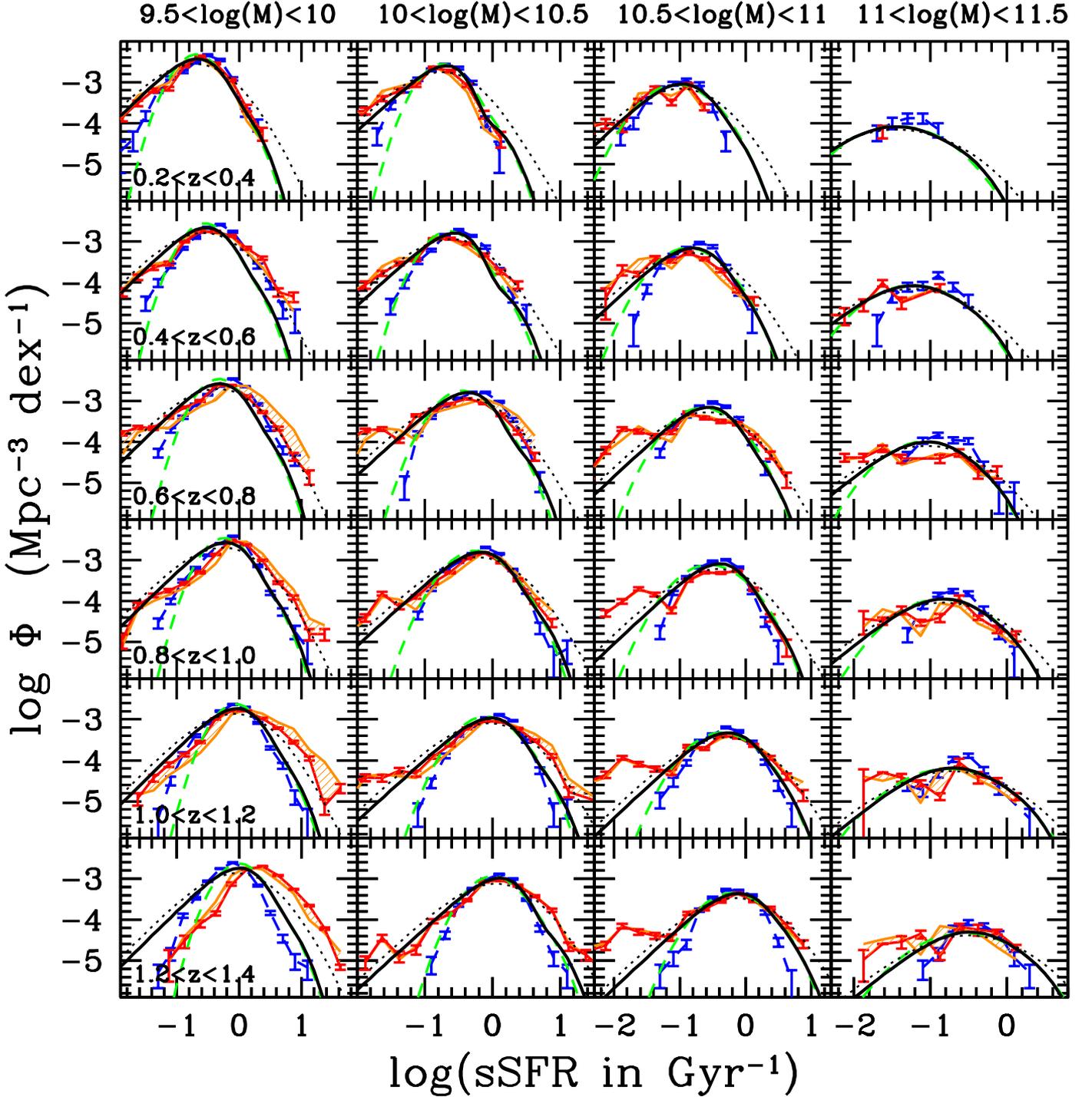} 
\caption{$s\SFR$ functions per redshift bin from $0.2<z<0.4$ to
  $1.2<z<1.4$ (from the top to the bottom rows) and per stellar mass
  bin from $9.5<log(\Ms)<10$ to $11<log(\Ms)<11.5$ (from the left to
  the right columns). The black solid lines and green dashed lines
  correspond to the best-fit $s\SFR_{UV+IR}$ functions assuming a
  double-exponential and a log-normal profile, respectively (as shown
  in Fig.\ref{allLF}). The dotted lines correspond to the same
  function convolved with a Gaussian having $\sigma=0.3$ dex to mimic
  the expected uncertainties on $\SFR_{SED}$. The $s\SFR_{NRK}$
  functions are shown with blue error bars and dashed lines. They are
  derived using an optical tracer of the $\SFR$ developed by Arnouts
  et al. (2013). The red and orange lines are obtained using
  $\SFR_{SED}$ with and without a correction for possible biases in
  $\SFR_{SED}$.
  \label{LFtracer}}
\end{figure*}

\begin{figure*}[htb!]
\includegraphics[width=19.cm]{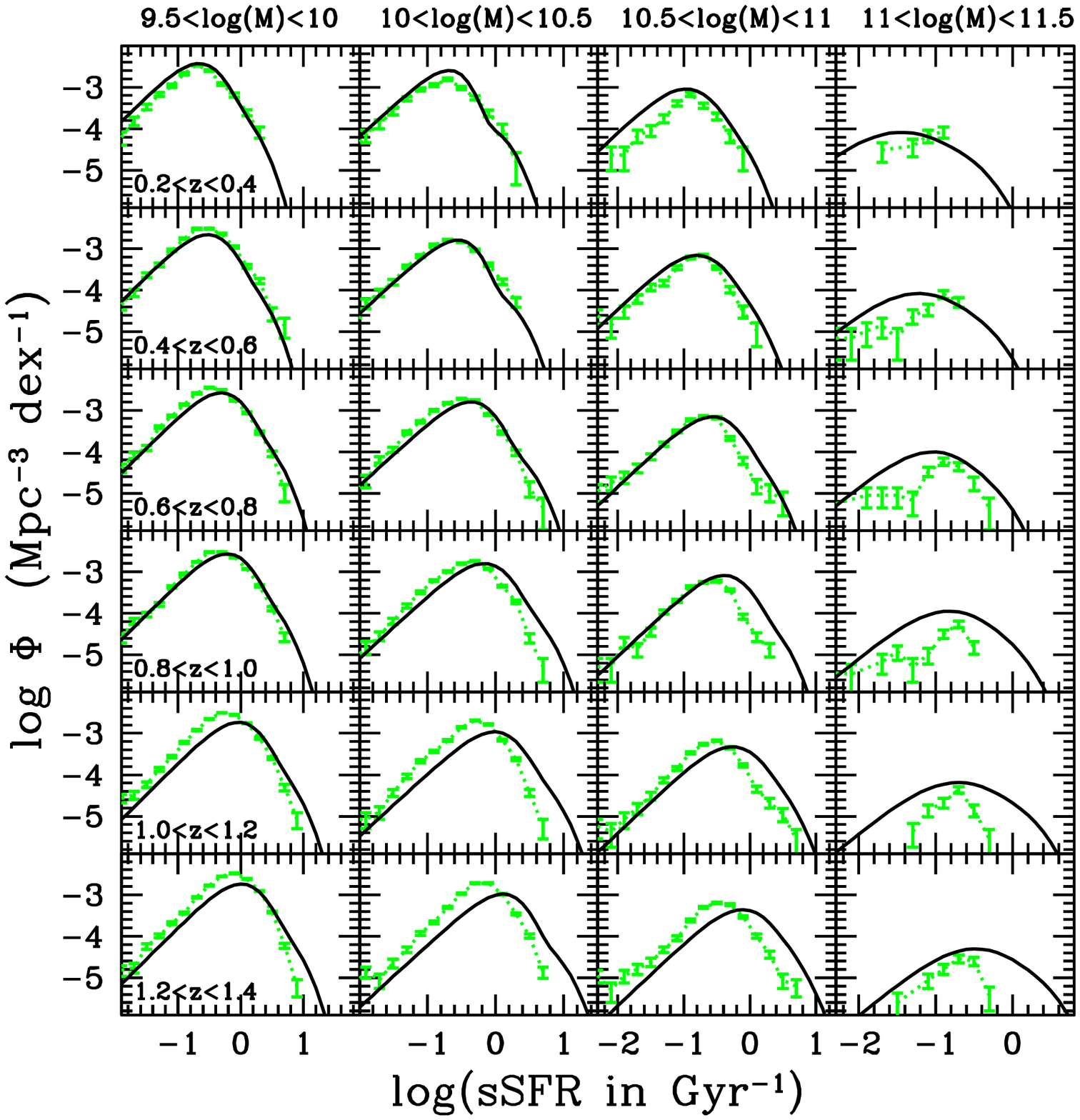} 
\caption{$s\SFR$ functions per redshift bin from $0.2<z<0.4$ to
  $1.2<z<1.4$ (from the top to the bottom rows) and per stellar mass
  bin from $9.5<log(\Ms)<10$ to $11<log(\Ms)<11.5$ (from the left to
  the right columns). The black solid lines correspond to the best-fit
  of the $s\SFR_{UV+IR}$ function with a double-exponential
  profile. The green dotted line corresponds to the predictions of the
  semi-analytical model.
  \label{LFIR}}
\end{figure*}

\subsection{Broadening of the $s\SFR$ function}\label{broad}

Figure \ref{sigma} shows the evolution of $\sigma$ as a function of
redshift, in the case of a log-normal fit. We find that $\sigma$
increases with mass. We also find that $\sigma$ is consistent with 
being constant with redshift at $z<1.4$.

Since $log(s\SFR)$ is not linearly proportional to $log(\Ms)$, it
could create an artificial broadening of the $s\SFR$ function,
especially if a large redshift range is considered. In order to test
this effect, we compute the sSFR functions in smaller mass bins of
$\Delta(log \Ms)=0.2$ rather than $\Delta(log \Ms)=0.5$. We still find
that $\sigma$ is close to 0.3 at $10.0<log(\Ms)<10.2$ and 0.45 at
$11.0<log(M)<11.2$.

We use the $\sigma$ value from the log-normal function since this
value can be directly compared with other values from the literature.
Our value of $\sigma$ is higher than previous studies which converge
to an r.m.s. of 0.2 dex (e.g. Peng et al. 2010, Sargent et al. 2012,
Salmi et al. 2013, Speagle et al. 2014). Noeske et al. (2007a) find an
r.m.s. of 0.35 dex, before deconvolution, which is close to our value
for the intermediate mass range. Almost no study finds a scatter of
$>0.4$ dex as we get for $\Ms>10^{11}\Msol$ galaxies, except Salim et
al. (2007).

An attractive interpretation is that the different mass ranges covered
by each survey could explain the various r.m.s. measured in the
literature. However, Salim et al. (2007) and Whitaker et al. (2012)
find that the scatter of the main sequence decreases with $\Ms$, which
is at odds with our result. Moreover, Lee et al. (2015) find an
r.m.s. of 0.35 dex almost constant with the $\Ms$ using similar data.

If we measure the r.m.s. of our own $\Ms-s\SFR$ scatter diagram
(Fig.\ref{masssfr}), we obtain an r.m.s. below 0.25 dex at
$log(\Ms)=11-11.5$, much lower than the $\sigma$ measured using the
$s\SFR$ function. One interpretation is that the dynamical $s\SFR$
range covered by the data is not sufficiently large to correctly
estimate the r.m.s. from a scatter diagram. We demonstrate this effect
with a simulated catalogue in \S\ref{SAM}. The advantage of using the
$s\SFR$ distribution is to extrapolate the shape of the function over
the full $s\SFR$ range, even if the data span a restricted $s\SFR$
range.

\subsection{Shape of the combined $s\SFR$ functions}\label{z0}

We correct for the redshift evolution of the $s\SFR$ and we combine
all the measurements at $z=0$. Figure \ref{LFz0} shows the combined
1/V$_{\rm max}$ data per stellar mass bin. The shape of the $s\SFR$
distribution appears invariant with time: the dispersion between the
data points is around 0.1 dex as shown in the insets of
Fig.\ref{LFz0}. We observe that the double-exponential and the
log-normal profiles provide a good fit of the combined data (solid red
line and blue dashed line, respectively). Even so, the $\chi^2$ values
are smaller for double-exponential fit at $10<log(\Ms)<11$.

As shown in Fig.\ref{sigma}, the broadening of the $s\SFR$ function is
also visible in the combined $s\SFR$ functions. If we let the fraction
of starbursts\footnote{The fraction of starburst is defined as the
  ratio between the integral of the log-normal distribution associated
  with the starburst component and the integral of the main-sequence
  contribution (log-normal or double-exponential). This contribution
  was set at 3\% when we fit individual redshift and mass bins.}  vary
while we fit the combined data, we obtain that the fraction of
starbursts is consistent with 0 at $log(\Ms)>10.5$, but the
uncertainties are consistent with a contribution of 1\%. At lower
mass, we find a fraction of starbursts of $2\pm1$\% and $4\pm 1$\%
assuming a double-exponential profile at $log(\Ms/\Msol)=10-10.5$ and
$9-9.5$. The associated uncertainties are underestimated since all the
parameters describing the shape of the starburst contribution are set,
except the normalization. There is a hint that the fraction of
starburst increases at low masses. However, we would need a survey
covering a larger volume to cover the high $s\SFR$ range of the
distribution, since massive starbursting galaxies are rare galaxies.

\section{The sSFR function using other $\SFR$ tracers}\label{opt}

When using the $\SFR_{UV+IR}$ tracer, the density of star-forming
galaxies below the $s\SFR$ peak relies on the extrapolation of the
best-fit profile (double-exponential or log-normal). The UV and
optical $\SFR$ tracers allow us to cover the full $s\SFR$ range and
could bring some information at low $s\SFR$. In this section, we
derive the $s\SFR$ function using $\SFR$ tracers based on the stellar
emissivity only (without using the IR data), as shown in
Fig.\ref{LFtracer}.

We estimate the $\SFR$ from the SED fitting procedure (the same method
as for the stellar mass, see \S\ref{data}). When comparing
$\SFR_{SED}$ and $\SFR_{UV+IR}$, we find some bias reaching 0.25 dex
and a scatter between 0.25 and 0.35 dex (the scatter increases both
with the mass and the redshift). The comparison between our reference
$s\SFR_{UV+IR}$ functions (black solid lines) and the $s\SFR_{SED}$
functions (red solid lines) shows that the $s\SFR_{SED}$ functions are
much flatter \footnote{Lower density at the density peak, and larger
  density at high/low $s\SFR$ than the reference $s\SFR_{UV+IR}$
  function.}. We convolve the reference $s\SFR_{UV+IR}$ functions by a
Gaussian function with $\sigma=0.3$ dex (black dotted lines) to mimic
the expected uncertainties. After this convolution, the agreement
between the convolved $s\SFR_{UV+IR}$ function (dotted lines) and the
$s\SFR_{SED}$ function is better. The $s\SFR_{SED}$ function would
favor a fit with a double-exponential profile. At high mass $\Ms >
10^{10.5}\Msol$, the density of low $s\SFR_{SED}$ galaxies even
exceeds what we expect from the double-exponential extrapolation.

Arnouts et al. (2013) have developed a new method for estimating the
$\SFR$ from optical data. This $\SFR$ - denoted $\SFR_{NRK}$ - is
estimated by correcting the UV intrinsic luminosity ${\cal L_{UV}}$ by
the infrared excess $IRX={\cal L_{IR}}/{\cal L_{UV}}$, directly
estimated from the position of the galaxy into the $NUV-R-K$ plane. We
use the parametrization of the $IRX$ from Arnouts et al. (2013),
slightly modified by Le Floc'h et al. (2014, in prep)\footnote{In the
  mass bin $9.5<log(\Ms)<10$, Le Floc'h et al. (2015, in prep) show
  that the IRX could be overestimated. Based on a stacking procedure
  using Herschel images, Le Floc'h et al. (2014, in prep) derive an
  additive term of $-0.35(z-0.25)$ to be added to the IRX at
  $9.5<log(\Ms)<10$.}. The dispersion between $\SFR_{NRK}$ and
$\SFR_{IR+UV}$ is around 0.15 dex (could reach 0.2 dex). Figure
\ref{LFtracer} shows the excellent agreement between the $s\SFR_{NRK}$
functions (blue open stars) and the $s\SFR_{IR+UV}$ functions (black
solid lines) at $\Ms < 10^{11}\Msol$.  The positions of the peak of
the $s\SFR$ functions are similar (within 0.2 dex).  Using the
$s\SFR_{NRK}$ tracer, we find a better agreement with the log-normal
profile since the $s\SFR_{NRK}$ density falls sharply in the lowest
$s\SFR$ bin. Still, the double-exponential profile is more appropriate
in some bins (for instance, in the redshift bin $0.2<z<0.4$). Given a
possible bias in $s\SFR_{NRK}$ at low $s\SFR$ (Arnouts et al. 2013)
leading to an underestimation of the $s\SFR_{NRK}$, we cannot
conclude that this sharp cut at low $s\SFR_{NRK}$ is real.

To summarize, given the large uncertainties affecting the UV and
optical $\SFR$ tracers, it is still challenging to constrain the low
$s\SFR$ end.

Finally, we note that our reference $\SFR$ tracer combining UV and IR
could be overestimated, since dust could be heated by the old stellar
populations (Utomo et al. 2014). This bias would mainly affect
galaxies with $s\SFR[Gyr^{-1}]<-1$. Correcting for such a bias would
modify the shape of our $s\SFR$ function: the slope of the low $s\SFR$
side of the $s\SFR$ function obtained with the double-exponential
profile would be even flatter than the observed one.

\begin{figure}[htb!]
\includegraphics[width=8.5cm]{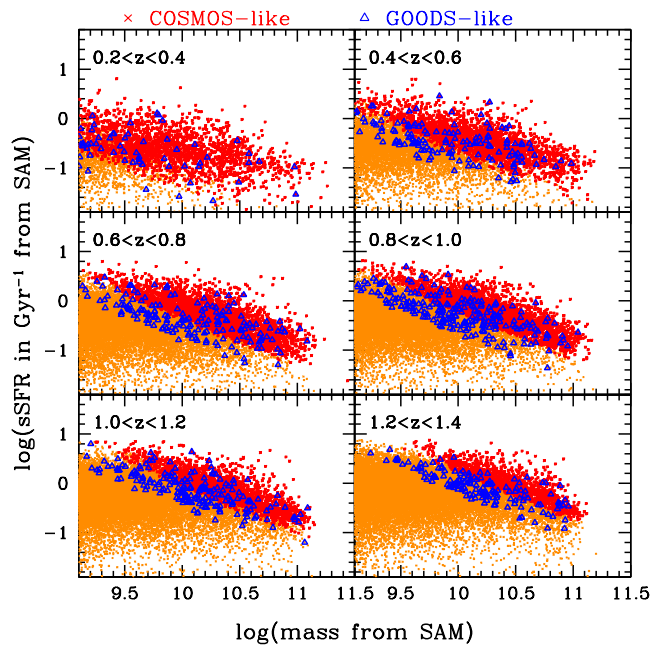}
\caption{$s\SFR$ as a function of the stellar mass using the
  prediction of the semi-analytical model. The orange points are the
  mass and the $s\SFR$ of the full simulated catalogue. The blue
  triangles and the red crosses correspond to a GOODS-like and a
  COSMOS-like survey, respectively.
\label{massssfr_simu}}
\end{figure}

\section{Comparison with a semi-analytical model}\label{SAM}

We now compare our results with the predictions of a semi-analytical
model. The mock catalogue is based on $\Lambda$CDM simulations from
Wang et al. (2008) and the galaxy properties were generated using the
galaxy formation model, as detailed in De Lucia \& Blaizot (2007) and
Wang \& White (2008). The light cone survey covers an area of
1.4$\times$1.4 deg$^2$ similar to COSMOS. The redshift, the $\SFR$ and
the stellar mass are available for all galaxies in the simulation, as
well as the observed magnitudes expected for these galaxies. We select
the star-forming galaxies using a criterion similar to our selection
in the $NUV-R-K$ plane.

We first test if we can reproduce the same selection effects as
discussed in \S\ref{scatter}. We apply a $K$-band selection in the
simulation similar to the ones applied in the data ($K<24$ and
$K<24.3$ for the COSMOS and GOODS surveys, respectively). The
selection at 24$\mu m$ creates an observational limit in the
redshift-$\SFR$ plane. We apply the same $\SFR$ limits in the
simulation as the ones established for the COSMOS and the GOODS
surveys. Finally, we select galaxies over an area of 1.5 $deg^2$ for
COSMOS and 138 $arcmin^2$ for GOODS.  The blue and red points in
Fig.\ref{massssfr_simu} show the distribution of the simulated sources
in the $\Ms-s\SFR$ plane for the GOODS-like and COSMOS-like surveys,
respectively. We reproduce exactly the same selection effect as the
ones discussed in \S\ref{scatter}.  The predicted COSMOS-like and
GOODS-like surveys do not cover the same area in the $\Ms-s\SFR$
plane. Even with the GOODS-like survey, the MIPS data are not
sufficiently deep to provide a representative sample of low-mass
galaxies in terms of $s\SFR$.

We also test that the width of the main sequence is not correctly
measured using simply the r.m.s. of the sample without taking into
account selection effects. For instance, the r.m.s. of the $s\SFR$
without any selection (orange points) is 0.38 dex at $0.8<z<1$ and
$9.5<log(\Ms)<10$, but we only measure an r.m.s. of 0.18 dex and 0.23
dex in the COSMOS-like and GOODS-like survey, respectively. It
illustrates the necessity of taking into account selection effects in
$\SFR$ limited surveys, as discussed in Rodighiero et al. (2014) and
Kelson (2014). In particular, any study looking at the evolution of
the $s\SFR$ with the mass would be biased.

Finally, we directly compute the predicted $s\SFR$ functions from the
simulated catalogue. A comparison with the $s\SFR$ functions predicted
by the models (green lines) and the observed ones is shown in
Fig.\ref{LFIR}. A qualitative comparison shows that the predicted
shape of the $s\SFR$ functions is similar to the observed one. A
parametrization with a double-exponential profile is perfectly
suitable for the simulation. In specific redshift and mass bins, the
agreement with the data is remarkable (e.g. $0.4<z<0.6$ and
$log(\Ms)<10.5$). The slope of the predicted $s\SFR$ function is in
excellent agreement with the double-exponential profile. The predicted
density of low $s\SFR$ star-forming galaxies exceeds the density
extrapolated from the log-normal profile. Therefore, an extrapolation
with the double-exponential profile is more natural than a log-normal
profile on the theoretical point of view.

The agreement between the predicted and observed $s\SFR$ functions
breaks down for galaxies more massive than $log(\Ms)>10.5$, but also
at $z>1$. As a global trend, the galaxies with the largest $s\SFR$ are
missed in the simulation (e.g. $0.8<z<1$ and $log(\Ms)>10.5$). At
$z>1$, the predicted distribution is shifted at lower $s\SFR$ in
comparison to the data.  We will discuss in \S\ref{SMIR} the predicted
evolution of the median $s\SFR$ compared to the observed one.

Finally, we note that we do not use the most recent SAMs. We keep the
same SAM as in our previous works in COSMOS. For instance, we use the
same SAM to compare predicted and observed $\MF$ as in Ilbert et
al. (2013). However, more detailed comparisons with recent numerical
simulations will be necessary in the future.

\begin{figure}[htb!]
\centering \includegraphics[width=9cm]{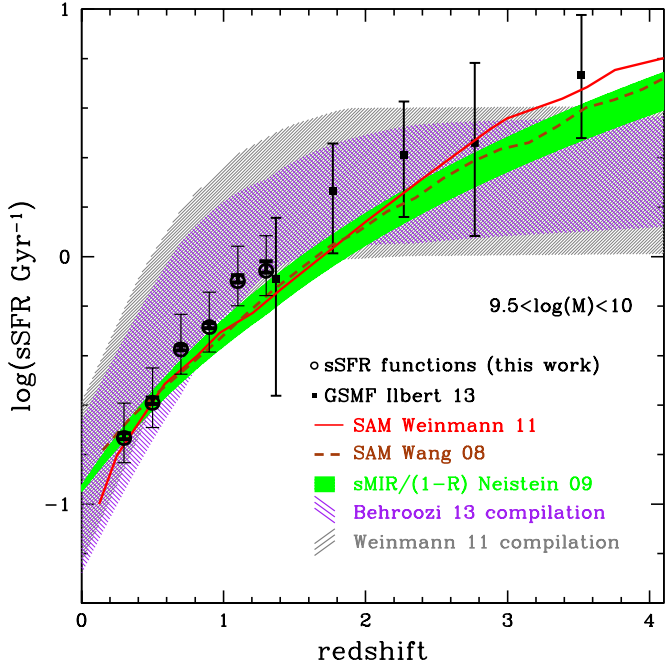}
\caption{Evolution of the median $s\SFR$ derived from the $s\SFR$
  functions at $9.5<log(\Ms/\Msol)<10$ (open black circles). The
  statistical uncertainties on the median $s\SFR$ are within the
  symbols. Systematic uncertainties ($\pm$0.1 dex in stellar mass and
  $+0.1$ dex in $\SFR$) are indicated with thin error bars. The
  $s\SFR$ derived indirectly from the UltraVISTA mass functions are
  indicated with filled black squares. The gray and purple shaded areas
  correspond to the data compilations from Weinmann et al. (2011) and
  Behroozi et al. (2013), respectively. The prediction of the SAM from
  Weinmann et al. (2011) and Wang et al. (2008) are shown with the red
  and brown lines. The green shaded area corresponds to the analytical
  relation from Neistein \& Dekel (2008) to describe the $sMIR$
  evolution, corrected for the mass loss as discussed in Appendix A.
  \label{sSFRweinmann}}
\end{figure}

\begin{figure}[htb!]
\includegraphics[width=9cm]{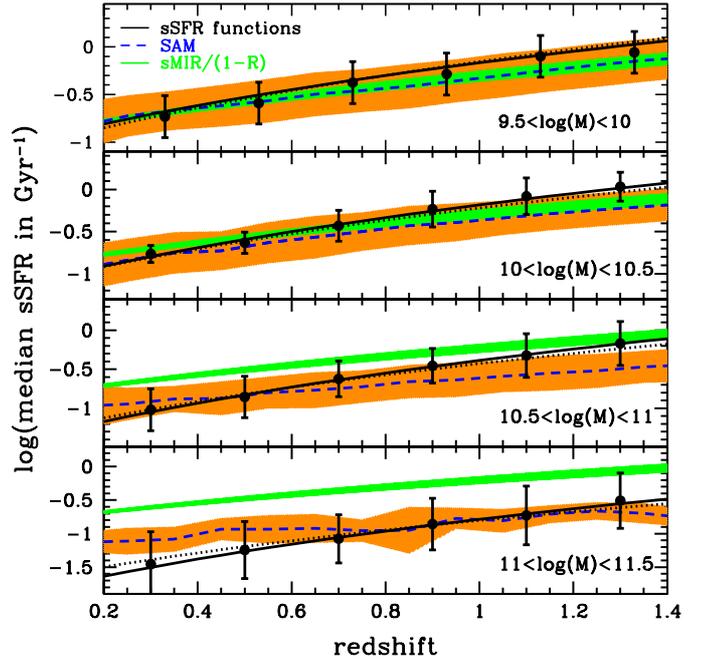}
\caption{Evolution of the median $s\SFR$ as a function of
  redshift. Each panel corresponds to a stellar mass bin.  The blue
  dashed lines correspond to the median $s\SFR$ expected from the
  semi-analytical model. The orange area is derived by measuring the
  r.m.s. of the $s\SFR$ in the semi-analytical model. The solid
  circles correspond to the median $s\SFR$ measured in this work. The
  vertical error bars indicate the $\sigma$ derived from the fit with a
  log-normal function (i.e., the intrinsic scatter of the $\Ms-s\SFR$
  relation). The solid (dotted) lines correspond to the fit over the
  data using Eq.\ref{ssfrEvol} assuming that $b$ does (does not)
  depend on the mass. The green shaded area corresponds to the
  analytical relation from Neistein \& Dekel (2008) to describe the
  $sMIR$ evolution, corrected for the mass loss as discussed in
  Appendix A.
  \label{sSFRsimu}}
\end{figure}

\section{Discussion}\label{discussion}

In this section, we discuss our main results: 1) the mass-dependent
increase in the $s\SFR$ with redshift; 2) the decrease in $log(s\SFR)$
as $-0.17\Ms$; and 3) the broadening of the $s\SFR$ function with
mass. We discuss here the numerous complex processes that can reduce
the $s\SFR$ as the stellar mass increases, from the hot halo quenching
mode to secular evolution of galaxy disks. The diversity of these
processes could explain the broadening of the $s\SFR$ functions with
mass, and their complexity could reduce the ability of the SAM to
reproduce the $s\SFR$ evolution for the most massive galaxies.

\subsection{Increase in $s\SFR$ with redshift: Link with the
  cosmological accretion rate}\label{SMIR}

We compare here the evolution of the median $s\SFR$ with the specific
mass increase rate $sMIR_{DM}$ ($\dot{M}_{H}/M_H$ following Lilly et
al. 2013) and with the predictions of semi-analytical models.

Assuming the gas inflow rate is driven by the cosmological accretion
rate of the dark matter structures, we expect that the $s\SFR$ follows
the evolution of the $sMIR_{DM}$ (in the following, we implicitly
divide the $sMIR_{DM}$ by $1-R$ with $R$ the return fraction, as
discussed in Appendix A). In simple models in which galaxies reach a
quasi-steady state (Bouch\'e et al. 2010, Lilly et al. 2013), the
evolution of the $s\SFR$ is coupled with the evolution of the
$sMIR_{DM}$. Based on N-body simulations and extended Press-Schechter
formalism, Neistein \& Dekel (2008) show that $sMIR_{DM}$ evolves as
$\propto 0.047
(M_H/10^{12}\Msol)^{0.15}\times(1+z+0.1(1+z)^{-1.25})^{2.5}$, which
could explain why the $s\SFR$ increases with redshift. The green
shaded areas in Fig.\ref{sSFRweinmann} and Fig.\ref{sSFRsimu} show the
evolution of the $sMIR_{DM}$, after having determined the value of
$M_H$ using the stellar-to-halo mass ratio from Coupon et al. (2015).

We first discuss the sample of low-mass galaxies at
$9.5<log(\Ms)<10$. In Fig.\ref{sSFRweinmann}, we show the evolution of
$sMIR_{DM}$ and we add the $s\SFR$ evolution predicted by the SAM from
Weinmann et al. (2011) (red solid line) and Wang et al. (2008) (brown
dashed line). In this mass range, the evolution of the $s\SFR$
predicted by the SAM follows closely the evolution of $sMIR_{DM}$
(Weinmann et al. 2011). This statement is also true even with the
latest results from the hydrodynamical Illustris simulation (Sparre et
al. 2014). We also show in Fig.\ref{sSFRweinmann} the data
compilations from Weinmann et al. (2011) and from Behroozi et
al. (2013) from various measurements available in the literature (gray
and magenta shaded areas). As discussed by Weinmann et al. (2011), the
observed $s\SFR$ from the literature are well above the predictions of
the SAM at $z<1.5$. We add in Fig.\ref{sSFRweinmann} our own
measurements of the median $s\SFR$. Our values are located in the
lowest part of the Weinmann et al. (2011) and from Behroozi et
al. (2013) compilations. Therefore, we find a much better agreement
between the observed and theoretical evolution of the $s\SFR$, as
expected if the gas feeding is directly driven by the cosmological
accretion rate. There are several reasons for the difference with
previous results: 1) we take into account selection effects that lead
to a lower median $s\SFR$ value than the ones obtained directly from a
$\SFR$ limited survey; 2) the previous compilations do not
differentiate between median and average $s\SFR$ which could modify
the $s\SFR$ values by 0.2 dex; or 3) a systematic uncertainty of $-0.1$
dex could affect our $\SFR$ measurements as discussed in
\S\ref{data}. Error bars in Fig.\ref{sSFRweinmann} include these
systematic uncertainties, as well as a possible $\pm0.1$ dex
systematic uncertainty on the stellar mass.

While the $s\SFR$ evolution matches the SAM predictions and follows
the $sMIR_{DM}$ evolution for our low-mass sample, this agreement
breaks down at higher masses. Figure \ref{sSFRsimu} shows the
evolution of the median $s\SFR$ predicted by the Wang et al. (2008)
model as well as the evolution of the $sMIR_{DM}$ in several stellar
mass bins. We first note that the evolution of $sMIR_{DM}$ no longer
corresponds to the evolution of the $s\SFR$ in the SAM. Indeed, AGN
feedback is included in the SAM in order to quench the star formation
in massive halos (e.g. Croton et al. 2006, Cattaneo et al. 2006, Wang
et al. 2008). While these recipes are sufficient to recover a broad
agreement with the observed $s\SFR$, we obtain that the median $s\SFR$
evolves faster in our data than in the SAM of Wang et al. (2008). In
the data, $b$ varies from 2.9 to 3.8 from low-mass to high-mass
galaxies. We find the reverse trend in the simulation.  The simulation
predicts that $b$ decreases with mass: $b=$2.3, 2.1, 1.9, 1.5 at
$log(\Ms)=9.5-10$ dex, $10-10.5$, $10.5-11$ and $11-11.5$,
respectively. We also observe that the width of the $s\SFR$ function
is smaller in the model than in the data for the massive galaxies. The
simulated scatter of the $s\SFR$ distribution is 0.22 dex at $\Ms<10
^{11}\Msol$ but reaches 0.16 dex for the most massive
galaxies. Therefore, the trend is the reverse of the observed one.

In Ilbert et al. (2013), the low-mass end of star-forming $\MF$ is
accurately reproduced by the SAM model of Wang et al. (2008) while the
model under-predicts the density of massive star-forming galaxies
(their Fig.14). Here, we also show that the evolution of the $s\SFR$
with redshift is in agreement with the evolution predicted by the SAMs
for low-mass galaxies, but complex physical processes that could
affect the SFH in massive galaxies, such as quenching or secular
evolution, need to be modeled more accurately. In particular, galaxies with
the highest $s\SFR$ are missing in the simulation at $z\sim 1$ as
shown in \S\ref{SAM}.

\subsection {Gradual decline of the $s\SFR$ with the mass: Quenching processes and/or lower efficiency of the star formation}\label{quenching}

One of our main results is that the full $s\SFR$ distribution is
shifted toward lower $s\SFR$ as the mass increases, with $log(s\SFR)
\propto -0.17\Ms$. We discuss here possible mechanisms that could
create such a dependency on the stellar mass.

\subsubsection{Quenching processes}

A first hypothesis is that all massive galaxies are on their way to
being quenched and we observe galaxies transitioning toward an even
lower $s\SFR$.

In some scenarios, such as Hopkins et al. (2008), a major merger could
trigger a burst of star formation and then quench a galaxy in less
than 0.3 Gyr. This quenching process cannot be ongoing for all massive
star forming galaxies simultaneously: we would observe the density of
massive star-forming galaxies dropping rapidly with time, while the
high-mass end of the $\MF$ does not evolve significantly at $z<1$
(e.g. Arnouts et al. 2007, Ilbert et al. 2010, Boissier et
al. 2010). With such a short quenching timescale, star-forming
galaxies would be removed almost instantaneously from our considered
sample.

Galaxy could also be quenched by an exhaustion of the cold gas supply
as the DM halos grow. For instance, cold accretion across filaments is
suppressed in massive halos at $z<2$ (Dekel et al. 2009) which reduces
the supply of cold gas and then in the star formation. Hydrodynamical
simulations predict the formation of a virial shock in dark matter
halos with $M_H>10^{12}\Msol$. These massive halos can be maintained
``hot'' with the radio AGN feedback mode or extreme star formation
feedback (e.g., Croton et al. 2006, Cattaneo et al. 2006, Wang et
al. 2008). According to Gabor \& Dav\'e (2015), mass quenching and
environment quenching would be the consequence of the same process:
the starving of the galaxies falling in a halo more massive than
$10^{12}\Msol$. The simple model of Noeske et al. (2007b) reflects the
SFH in such gas exhaustion case. In Noeske et al. (2007b), the
decreases in the $s\SFR$ with mass is reproduced by assuming
exponentially declining SFH with $\tau$ having an inverse dependency
with mass - $\tau \propto 1/\M$. The ''stage'' model of Noeske et
al. (2007b) reproduces well the turn-over at high mass that we
observe. For a galaxy as massive as $log(\Ms)=11.3$ at $z=0.5$, this
model associates an exponentially declining SFH with a $\tau$ value as
large as 4 Gyr. Therefore, the bending of the $s\SFR$ with mass could
be explained by gas exhaustion over long timescales $>3-4Gyr$ ($\tau$
value decreases with stellar mass). Such timescales are longer than
the one usually adopted in simulations to quench star formation in hot
halos (typically $1.2\pm 0.5$ Gyr for Gabor \& Dav{\'e} 2015, private
communication). Therefore, if the quenching in the hot halo mode
explains the bending of the $s\SFR$ at high mass, it should occur on a
longer timescale than usually assumed in simulation. However, if the
quenching in the hot halo mode occurs with a timescale $<1Gyr$, this
process does not explain the bending of the $s\SFR$ with mass
(although this process could be perfectly relevant to explain the
formation of the quenched galaxies over time).

While these quenching processes are probably crucial to generating the
quiescent population, they do not likely explain the evolution of the
$s\SFR$ with mass in the star-forming population since they act on too
short a timescale.

\subsubsection{Declining efficiency of the star formation toward massive systems - Impact of the bulge}

A second possibility is that the efficiency in forming new stellar
populations declines slowly as the stellar mass increases, without
necessarily quenching. Kassin et al. (2012) show that the massive
galaxies are on average more kinetically settled at $0.2<z<1.2$. They
find a possible threshold around $10^{10.4} \Msol$ to move from a
disordered to settled disk. If we speculate that random motion in the
gas is going in the direction of a higher star formation efficiency,
it could explain a decrease in the $s\SFR$ above $10^{10.4} \Msol$.
Sheth et al. (2008) show that most massive spiral galaxies have a
higher fraction of bars, associated with a bulge and having redder
colors. They suggest that massive systems are more mature.

Abramson et al. (2014) show in the SDSS that the decrease in the
$s\SFR$ with stellar mass is explained by the increase in the
bulge-mass fraction with stellar mass. The bulge is less efficient
in forming stars, which explains a decrease in $s\SFR$ with mass. At
$0.5<z<2.5$, Lang et al. (2014) show that the mass fraction within the
bulge increases from 30\% in $10^{10}\Msol$ star-forming galaxies to
50\% in $10^{11}\Msol$ star-forming galaxies. Surprisingly, these
ratios remain consistent between $z\sim 1$ and $z\sim 2$. Therefore,
the mass contribution of the bulge to the total mass increases with
$\Ms$ at all redshifts. 

The presence of the bulge could lower the star formation efficiency.
In the local Universe, Saintonge et al. (2012) show that the depletion
timescale of molecular gas is longer when the galaxy is
bulge-dominated, pointing to a lower star formation efficiency when a
bulge is present. Using hydrodynamical simulation, Martig et
al. (2009) show that a bulge stabilizes the disk against fragmentation
and this process suppresses the formation of massive star-forming
clumps in the inner part of the galaxy. Genzel et al. (2014)
show that the Toomre parameter $Q$ increases at the galaxy center for
a sample of $z\sim 2$ massive disk galaxies, shutting off the
gravitational instability and reducing the star formation efficiency
in the inner part of the disk. Finally, F{\"o}rster-Schreiber et
al. (2014) show that the presence of AGN-driven massive outflows in
the nuclear region that are visible only for their most massive disk
galaxies at $z\sim 2$ ($\Ms>10^{11}\Msol$). Such outflows could clear
the inner region from the gas and suppress the star formation in the
bulge.

We note that the decline (or even the shut down) of the star formation
in the inner region of the galaxy does not imply a quenching of the
star formation in the entire galaxy. We take as an example the case of
the Milky Way (MW). Snaith et al. (2014) and van Dokkum et al. (2013)
analyze the SFH of the MW. For a lookback time of 6 Gyr, which
corresponds to $z\sim 0.7$, these studies expect a $\SFR$ below
3$\Msol/yr$ and $log(\Ms)=10.6$. From Fig.\ref{allLF}, the MW falls in
the declining part of the $s\SFR$ function with a $log(s\SFR
[Gyr^{-1}])=-1.1$ dex. Six Gyr later, the MW is not yet a quiescent
galaxy and still forming 1 $\Msol/yr$ (e.g. van Dokkum et
al. 2013). There is no reason why the MW should quench on a timescale
of a few Gyr. Therefore, these massive galaxies with a low $s\SFR$ are
not necessarily quenching but could simply be quietly forming stars
along cosmic time, as in the MW. A significant density of low $s\SFR$
star-forming galaxies is expected in the SAM (see \S\ref{SAM}), in
agreement with a double-exponential profile for the $s\SFR$
function. Unfortunately, the small dynamical $s\SFR$ range covered by
our $\SFR_{UV+IR}$ tracer in COSMOS and GOODS (see \S\ref{z0}), as
well as the large uncertainties within the $\SFR_{SED}$ and
$\SFR_{NRK}$ tracers (see \S\ref{opt}) do not allow us to definitively
come to a conclusion about the density of the low $s\SFR$ galaxies not
yet quenched.

Finally, we note that the bulge formation could be done through two
channels, through secular evolution and by major and minor mergers. In
the former case, a bulge could form along time under the action of a
bar (e.g. Perez et al. 2013), or through gravitational disk
instabilities with large star forming clumps moving inward
(e.g. Immeli et al. 2004, Bournaud et al. 2007, Genzel et al. 2008,
Bournaud et al. 2011, Perez et al. 2013). Our analysis provides useful
information on the SFH of the galaxy which evolves secularly. However,
if the bulge originates from galaxy mergers (e.g. Kauffmann \&
Haehnelt 2000, Martig et al. 2009), the stellar mass has not been
formed in situ which makes the ratio $\SFR/(M_{bulge}+M_{disk})$
difficult to interpret in term of SFH (Abramson et al. 2014). In
general, mergers would bring stellar mass created ex-situ, leading to
a stellar mass growth. If the $\SFR$ is not triggered to a higher value
during the merger, it could lead to a growth in mass and then a
decrease in the $s\SFR$ (Peng et al. 2014).

\begin{figure*}[htb!]
\centering \includegraphics[width=15cm]{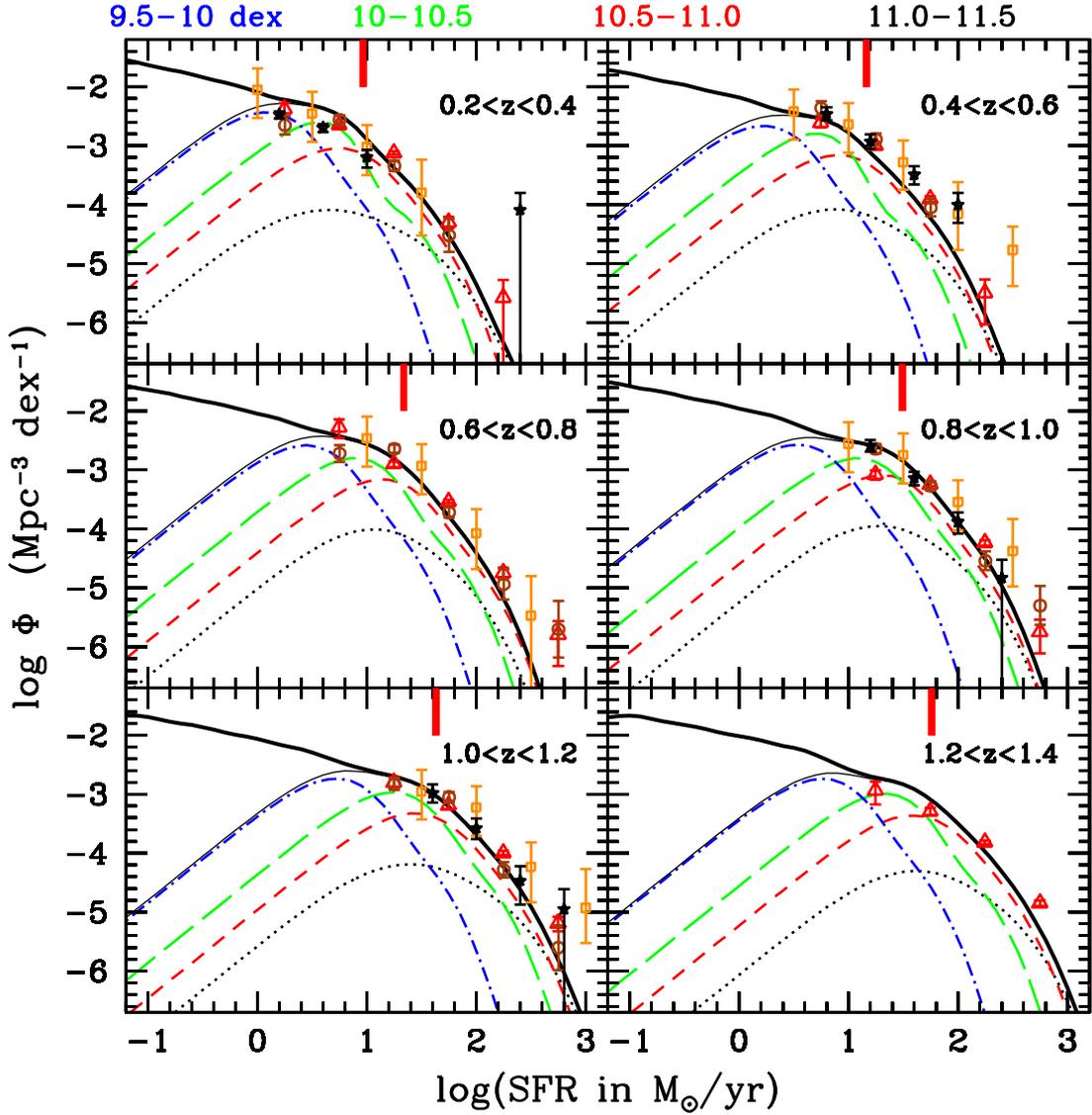}
\caption{$\SFR$ function per redshift bin (from $0.2<z<0.4$ in the top
  left panel to $1.2<z<1.4$ in the bottom right panel) and per stellar
  mass bin (dashed-dot blue: 9.5-10 dex, long-dashed green: 10-10.5,
  short-dashed red: 10.5-11 and dotted black 11-11.5). We show only
  the best-fit functions using a double-exponential profile and their
  sum corresponds to the light black solid line. The full $\SFR$
  function obtained by extrapolating the contribution of galaxies at
  $\Ms<10^{9.5}\Msol$ is shown with the thick solid black line. The
  points correspond to IR Luminosity Functions from the literature
  converted into $\SFR$ functions (red open triangles: Gruppioni et
  al. 2013; orange open squares: Le Floc'h et al. 2005; brown open
  circles: Rodiguiero et al. 2010; black filled stars: Magnelli et
  al. 2009). The thick red vertical lines indicate the location of
  $\SFR_{knee}$ ($log_{10}(\SFR_{knee})=$0.96, 1.16, 1.34, 1.49, 1.63,
  1.76 at z=0.2-0.4, 0.4-0.6, 0.6-0.8, 0.8-1, 1-1.2, 1.2-1.4)
\label{SFRfunction}}
\end{figure*}

\subsection{Broadening of the $\Ms-\SFR$ relation: Star formation stochasticity and diversity in SFHs}

As discussed in \S\ref{broad} and as shown in Fig.\ref{sigma}, we find
that the intrinsic scatter of the main sequence increases with
mass. In particular, the standard deviation found for the most massive
galaxies $11<log(\Ms)<11.5$ reaches $\sigma \sim 0.45$, which is well
above the values commonly found in the literature. In Appendix B, we
show that the intrinsic $s\SFR$ evolution and the criterion used to
select star-forming galaxies do not artificially create a broadening
of the $s\SFR$ function with the mass.

The intrinsic scatter of the $\Ms-\SFR$ relation indicates how tightly
the instantaneous star formation is determined by the past star
formation history of the galaxies. Numerous processes could scatter
the relation: the intrinsic scatter of the sMIR, galaxy mergers, the
variety of the possible SFHs, or the variation of the star formation
efficiency within the galaxy itself.

The dynamics of the gas and star content within a galaxy could create
$\SFR$ variations over million-year timescale. These variations create
a natural scatter around the $\Ms-\SFR$ relation (Dom{\'{\i}}nguez et
al. 2014). Hopkins et al. (2014) analyze the variability of eight
galaxies using hydrodynamical simulations and show that the
variability could easily reach 50\% for $\Ms^*$ galaxies when
considering a timescale of 20 millions years. In particular, SN
feedback has an important impact on this rapid variation. These
stochastic fluctuations result from variations in the star formation
efficiency over short timescales, generated mainly by the local impact
of SN feedback. Based on hydrodynamical simulations, Dom{\'{\i}}nguez
et al. (2014) show that these fluctuations generate an intrinsic
scatter in the $\Ms-\SFR$ relation reaching 0.5 dex for the dwarf
galaxies at $\Ms \sim 10^{7}\Msol$ which decreases at 0.2 dex for
intermediate mass galaxies at $\Ms\sim 10^{10}\Msol$. Since our lowest
mass range is $9.5<log(\Ms)<10$, we cannot detect such a decrease of
the scatter with the mass. Still, the intrinsic scatter that we
measure for our less massive galaxies $10^{9.5}<\Ms/\Msol< 10^{10}$
could be explained by the stochasticity of the star formation.

The intrinsic scatter in the $\Ms-\SFR$ relation induced by the star
formation stochasticity in individual galaxies decreases with the
stellar mass (Dom{\'{\i}}nguez et al. 2014). Therefore, this process
does not explain the increase in $\sigma$ that we find at $\Ms >
10^{10}\Msol$. We also do not expect the scatter of the sMIR to
increase with the halo mass. Indeed, we do not detect an increase in
the $s\SFR$ scatter in the SAM (see \S\ref{SMIR}).  Another
possibility is that the diversity in the possible SFH increases with
the mass. As discussed in \S\ref{quenching}, numerous processes could
affect the SFH and tend to reduce the $s\SFR$ as the mass
increases. Given the variety of these processes, and their possible
dependency on the halo mass and on the galaxy morphology (e.g. the
growth of a bulge), the impact on the SFH could vary
significantly. Therefore, the same processes could be simultaneously
responsible for the increasing diversity in the SFHs (i.e., the
scatter of the relation) and for the decrease in the $s\SFR$ with the
mass.

Kelson (2014) defines a statistical framework using the central limit
theorem to predict the width of the $\Ms-\SFR$ relation. In this
paper, the Hurst parameter $H$ (Kelson 2014 and references therein)
determines the behavior of the stochastic fluctuations in the $\SFR$.
For $H=0$, there is no stochastic fluctuation. For a value of $H=0.5$,
there is no covariance between the stochastic changes in $\SFR$, i.e.,
the expectation of the $\SFR$ at $t_{i+1}$ has the same value as the
$\SFR$ at $t_i$. Using the central limit theorem, Kelson (2014)
predicts a width of 0.3 dex for the $log(s\SFR)$ distribution when
$H=0.5$. For a value of $H=1$, i.e., the stochastic changes in $\SFR$
are strongly correlated with previous values (if the $\SFR$ decreases
at a given timestep, the $\SFR$ is more likely to decrease again the
following timestep). For a value of $H=1$, Kelson (2014) find that the
width of the distribution reaches 0.43 dex. If $H$ changes with the
mass, it could explain why we observe a variation of $\sigma$ from 0.3
to 0.45 dex between our two extreme considered bins. It implies that
stochastic changes in $\SFR$ are more correlated as the mass
increases, which could be seen as a larger diversity of SFH as the
mass increases. Kelson (2014) shows that the difference between the
median and the averaged $s\SFR$ could be used to establish the value
of $H$. We find that the differences between the median and the
average $s\SFR$ are around 0.1-0.15 dex in the lowest mass bin while
it increases at 0.2-0.25 dex in the highest mass bin (see Tables 1 and
2). Therefore, we measure that $H$ increases with stellar mass.

\begin{figure}[htb!]
\centering \includegraphics[width=9cm]{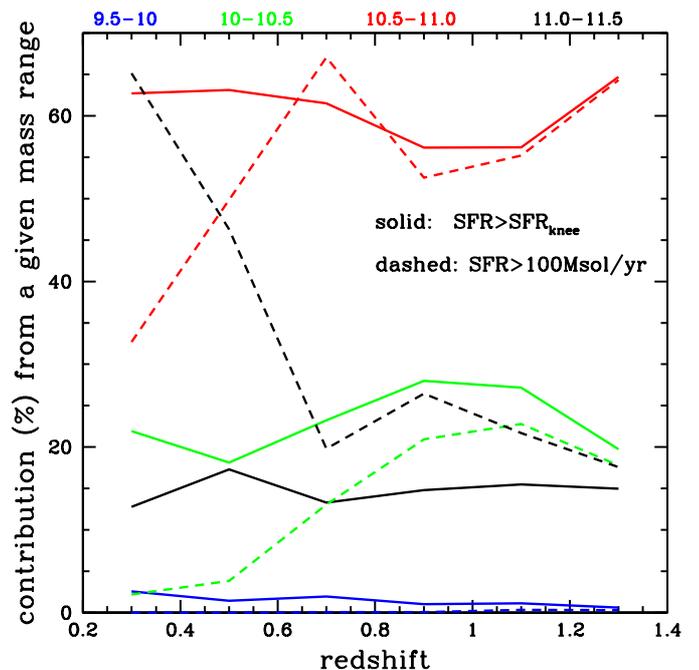}
\caption{Contribution in \% of a given population selected in stellar
  mass (blue: 9.5-10 dex, green: 10-10.5, red: 10.5-11 and black
  11-11.5) to the total $\SFR$ function integrated above a given
  $\SFR$. The dashed lines correspond to a $\SFR$ limit of
  100$\Msol/yr$ and the solid lines correspond to an evolving $\SFR$
  limit which is the ``knee'' of the $\SFR$ function.}
           \label{contributionSFRfunc} 
\end{figure}

\subsection{Interpreting the evolution of the $\SFR$ function and
  infrared luminosity functions}\label{SFRfunc}

In this section, we convert the $s\SFR$ functions into $\SFR$
functions and we discuss the evolution of the IR luminosity function
based on our results. This approach is complementary to Sargent et
al. (2012) and Bernhard et al. (2014), who combined the $\MF$ and an
universal $s\SFR$-distribution based on Rodighiero (2011) to interpret
the evolution of the $\SFR$ function.

The $\SFR$ function in a given mass bin is easily obtained by simply
adding the median of $log(\Ms)$ of the considered bin\footnote{9.75,
  10.25, 10.7 and 11.1 in the mass bin $log(\Ms)$=9.5-10, 10-10.5,
  10.5-11 and 11-11.5, respectively.} to the $s\SFR$ function. We sum
the $\SFR$ functions computed in several stellar mass bins to obtain
the total $\SFR$ function, as shown with a thin black line in
Fig.\ref{SFRfunction}.

Still, our data are limited to $\Ms>10^{9.5}\Msol$ and we need to
account for the contribution of the low-mass population when we derive
the global $\SFR$ function (thick black line in
Fig.\ref{SFRfunction}). Therefore, we assume that:
\begin{itemize} 
\item the density of star-forming galaxies (in log) evolves
  proportionally to $-0.4\log(\Ms)$, as derived from the $\MF$ of
  star-forming galaxies (e.g. Peng et al. 2010, Baldry et al. 2012,
  Ilbert et al. 2013, Tomczak et al. 2014)\footnote{$\alpha=-1.4$ in
    Peng et al. (2010) for a Schechter function or $\alpha_2=-1.5$ for
    a double-Schechter function in Baldry et al. (2012), Ilbert et
    al. (2013), Tomczak et al. (2014). When expressed per $d(log\Ms)$,
    the Schechter function has a slope evolving in $\alpha+1$, which
    explains why we adopt a factor of -0.4.};
\item the shape of the $s\SFR$ function at $log(\Ms)=9.5-10$ is
  conserved at lower mass. The width of the main sequence found in our
  lowest mass bin is similar to the one found in the deepest surveys
  (e.g. Whitaker et al. 2012);
\item our parametrization of the median $s\SFR$ evolution holds at
 $\Ms<10^{9.5}\Msol$. With our parametrization, $log(s\SFR)$ does not
  depend significantly on $\Ms$ at $\Ms<10^{9.5}\Msol$, in agreement
  with Lee et al. (2015) and Whitaker et al. (2014).
\end{itemize} 
We reconstruct the total $\SFR$ functions, as shown with the thick
solid lines in Fig.\ref{SFRfunction}. 

In Fig.\ref{SFRfunction}, we compare our total $\SFR$ function with
direct measurements from the literature. We convert the IR luminosity
functions into $\SFR$ functions following Kennicutt (1998). We find an
excellent agreement between our $\SFR$ functions and the ones derived
directly from the IR luminosity functions.

In the remaining part of this section, we use our results on the
$s\SFR$ to interpret the behavior of the IR luminosity function
discussed in the literature.

The total $\SFR$ function could be characterized as the combination of
two power laws (e.g. Magnelli et al. 2009), with a change in slope at
$\SFR_{knee}$. Assuming that the slope and the characteristic $\Ms^*$
of the star-forming $\MF$ do not evolve with time which is reasonable
at $z<1.4$ (e.g. Arnouts et al. 2007, Ilbert et al. 2010, Boissier et
al. 2010, Peng et al. 2010, Moustakas et al. 2013), the $\SFR_{knee}$
should evolve as the $s\SFR$. Using the $s\SFR$ evolution in
$3.2(1+z)$ found in \S\ref{evol}, we indeed reproduce the position of
the knee in Fig.\ref{SFRfunction} (vertical red thick line at top of
each panel).  By fitting the evolution of the knee of IR LFs, Le
Floc'h et al. (2005), Magnelli et al. (2011), and Magnelli et
al. (2013) find a consistent evolution of $3.2^{+0.7}_{-0.2}(1+z)$,
$3.5^{+0.5}_{-0.5}(1+z)$, and $3.8^{+0.6}_{-0.6}(1+z)$, respectively.

Figure \ref{contributionSFRfunc} shows the contribution of the
galaxies of a given stellar mass range to the full star-forming
population above a given $\SFR$ threshold.  If we use $\SFR_{knee}$ as
the $\SFR$ threshold, we obtain that the contribution of a given
stellar mass range remains stationary over the full redshift range
(solid lines). Figure \ref{contributionSFRfunc} also shows that
galaxies at $\Ms>10^{10}\Msol$ dominate the $\SFR$ function above
$\SFR_{knee}$, while galaxies with $\Ms<10^{10}\Msol$ contribute to
less than 5\%.

Several studies tried to reconcile the fact that the $\MF$ is
accuretely represented by a Schechter function while the IR luminosity
function is better represented by a double-exponential (e.g. Sargent
et al. 2012). We propose here the following interpretation: the $\SFR$
function could be seen as the $\MF$ convolved with the $s\SFR$
function. The density of star-forming galaxies drops above $\Ms^*$ in
the $\MF$, and the contribution of $\Ms>10^{11}\Msol$ galaxies stays
below 20\%. The high star-forming end is dominated by galaxies around
$\Ms^*$. The $\SFR$ of the galaxies around $\Ms^*$ will be spread
following their distribution in $s\SFR$. Therefore, the shape of the
$\SFR$ function above $\SFR^*$ is driven by the width of the $s\SFR$
function of $\Ms^*$ galaxies.

Finally, the faint-end slope of the $\SFR$ functions should also be a
power-law with the same slope as the star-forming $\MF$ if the $s\SFR$
does not depend on the mass at $\Ms<10^{9.5}\Msol$ (the term depending
on the mass in Eq.\ref{ssfrEvol} becomes negligible). Gruppioni et
al. (2013) find a slope of $-0.2$ and Magnelli et al. (2009) a slope
of $-0.6$ for the IR luminosity functions, while we would expect a
value around $-0.4$ from the $\MF$ (e.g., Peng et al. 2010, Baldry et
al. 2012, Ilbert et al. 2013, Tomczak et al. 2014).

\begin{table*}[htb!]
\begin{center}
\begin{tabular}{lc c c c c c  c c} \hline

                       &               &           &                   &                                &   &  median  &  average  & \\ 
                       &               &  N COSMOS         &                   &  $log(s\SFR^*)$           &   $\Phi^*$  &  $log(s\SFR)$ &  $log(s\SFR)$ & $\chi^2$ \\ 
$\Ms$ bin &    z-bin   &  + GOODS     & $\sigma$   &  (Gyr$^{-1}$)              &  ($10^{-3} Mpc^{-3}$) &  (Gyr$^{-1}$)   &  (Gyr$^{-1}$)   &     \vspace{0.2cm} \\ \hline
\hline
9.5-10 
 & combined &  & 0.217$^{{\rm +0.016}}_{{\rm -0.016}}$ & -1.028$^{{\rm +0.069}}_{{\rm -0.067}}$ & 2.912$^{{\rm +0.583}}_{{\rm -0.464}}$ & -1.069$^{{\rm +0.012}}_{{\rm -0.012}}$ & -0.940$^{{\rm +0.007}}_{{\rm -0.008}}$   & 154.021  \\
 & 0.2-0.4 & 1080+0  & 0.22 & -0.704$^{{\rm +0.012}}_{{\rm -0.013}}$ & 9.377$^{{\rm +0.222}}_{{\rm -0.219}}$ & -0.733$^{{\rm +0.012}}_{{\rm -0.012}}$ & -0.602$^{{\rm +0.012}}_{{\rm -0.012}}$   & 29.228  \\  
 & 0.4-0.6 & 643 +45 & 0.22 & -0.562$^{{\rm +0.014}}_{{\rm -0.014}}$ & 5.520$^{{\rm +0.118}}_{{\rm -0.117}}$ & -0.591$^{{\rm +0.014}}_{{\rm -0.012}}$ & -0.460$^{{\rm +0.014}}_{{\rm -0.014}}$   & 9.621  \\  
 & 0.6-0.8 & 1009+92 & 0.22 & -0.347$^{{\rm +0.010}}_{{\rm -0.011}}$ & 6.624$^{{\rm +0.105}}_{{\rm -0.105}}$ & -0.375$^{{\rm +0.010}}_{{\rm -0.010}}$ & -0.245$^{{\rm +0.010}}_{{\rm -0.010}}$   & 14.771  \\ 
 & 0.8-1.0 & 1111+61 & 0.22 & -0.258$^{{\rm +0.010}}_{{\rm -0.010}}$ & 6.815$^{{\rm +0.092}}_{{\rm -0.092}}$ & -0.285$^{{\rm +0.010}}_{{\rm -0.010}}$ & -0.156$^{{\rm +0.010}}_{{\rm -0.010}}$   & 21.068  \\ 
 & 1.0-1.2 & 546 +60 & 0.22 & -0.072$^{{\rm +0.015}}_{{\rm -0.016}}$ & 4.660$^{{\rm +0.070}}_{{\rm -0.070}}$ & -0.099$^{{\rm +0.014}}_{{\rm -0.016}}$ & 0.030$^{{\rm +0.015}}_{{\rm -0.015}}$   & 31.262  \\  
 & 1.2-1.4 & 180 +29 & 0.22 & -0.028$^{{\rm +0.023}}_{{\rm -0.026}}$ & 4.602$^{{\rm +0.064}}_{{\rm -0.064}}$ & -0.057$^{{\rm +0.024}}_{{\rm -0.022}}$ & 0.073$^{{\rm +0.024}}_{{\rm -0.024}}$   & 9.320  \\   

\hline
10-10.5  
 & combined &  & 0.184$^{{\rm +0.010}}_{{\rm -0.011}}$ & -1.018$^{{\rm +0.046}}_{{\rm -0.045}}$ & 4.183$^{{\rm +0.563}}_{{\rm -0.473}}$ & -1.177$^{{\rm +0.006}}_{{\rm -0.008}}$ & -1.060$^{{\rm +0.005}}_{{\rm -0.006}}$   & 324.531  \\
 & 0.2-0.4 & 991 +0  & 0.100$^{{\rm +0.014}}_{{\rm -0.000}}$ & -0.230$^{{\rm +0.012}}_{{\rm -0.078}}$ & 28.446$^{{\rm +0.706}}_{{\rm -6.068}}$ & -0.765$^{{\rm +0.012}}_{{\rm -0.012}}$ & -0.671$^{{\rm +0.012}}_{{\rm -0.012}}$   & 18.421  \\
 & 0.4-0.6 & 1193+31 & 0.126$^{{\rm +0.025}}_{{\rm -0.026}}$ & -0.229$^{{\rm +0.145}}_{{\rm -0.124}}$ & 12.144$^{{\rm +6.500}}_{{\rm -3.620}}$ & -0.631$^{{\rm +0.016}}_{{\rm -0.018}}$ & -0.532$^{{\rm +0.013}}_{{\rm -0.014}}$   & 41.747  \\
 & 0.6-0.8 & 1698+44 & 0.183$^{{\rm +0.021}}_{{\rm -0.022}}$ & -0.270$^{{\rm +0.097}}_{{\rm -0.090}}$ & 5.883$^{{\rm +1.819}}_{{\rm -1.264}}$ & -0.431$^{{\rm +0.016}}_{{\rm -0.016}}$ & -0.314$^{{\rm +0.012}}_{{\rm -0.012}}$   & 15.312  \\
 & 0.8-1.0 & 2239+40 & 0.213$^{{\rm +0.018}}_{{\rm -0.019}}$ & -0.180$^{{\rm +0.078}}_{{\rm -0.074}}$ & 4.277$^{{\rm +0.970}}_{{\rm -0.735}}$ & -0.233$^{{\rm +0.014}}_{{\rm -0.014}}$ & -0.105$^{{\rm +0.011}}_{{\rm -0.011}}$   & 41.264  \\
 & 1.0-1.2 & 1570+71 & 0.217$^{{\rm +0.021}}_{{\rm -0.022}}$ & -0.044$^{{\rm +0.092}}_{{\rm -0.087}}$ & 2.807$^{{\rm +0.757}}_{{\rm -0.549}}$ & -0.081$^{{\rm +0.018}}_{{\rm -0.018}}$ & 0.048$^{{\rm +0.013}}_{{\rm -0.013}}$   & 40.360  \\
 & 1.2-1.4 & 828 +34 & 0.170$^{{\rm +0.087}}_{{\rm -0.067}}$ & 0.245$^{{\rm +0.339}}_{{\rm -0.402}}$ & 4.428$^{{\rm +7.457}}_{{\rm -2.210}}$ & 0.033$^{{\rm +0.578}}_{{\rm -0.134}}$ & 0.146$^{{\rm +0.578}}_{{\rm -0.133}}$   & 8.663  \\

\hline
10.5-11  
 & combined &  & 0.274$^{{\rm +0.012}}_{{\rm -0.012}}$ & -1.605$^{{\rm +0.049}}_{{\rm -0.049}}$ & 1.566$^{{\rm +0.198}}_{{\rm -0.173}}$ & -1.453$^{{\rm +0.012}}_{{\rm -0.010}}$ & -1.300$^{{\rm +0.007}}_{{\rm -0.007}}$   & 157.307  \\
 & 0.2-0.4 & 539 +0  & 0.269$^{{\rm +0.039}}_{{\rm -0.042}}$ & -1.156$^{{\rm +0.160}}_{{\rm -0.149}}$ & 1.474$^{{\rm +0.773}}_{{\rm -0.460}}$ & -1.019$^{{\rm +0.030}}_{{\rm -0.030}}$ & -0.869$^{{\rm +0.023}}_{{\rm -0.024}}$   & 4.849  \\
 & 0.4-0.6 & 697 +11 & 0.265$^{{\rm +0.035}}_{{\rm -0.037}}$ & -0.981$^{{\rm +0.141}}_{{\rm -0.136}}$ & 1.165$^{{\rm +0.517}}_{{\rm -0.331}}$ & -0.855$^{{\rm +0.024}}_{{\rm -0.028}}$ & -0.708$^{{\rm +0.020}}_{{\rm -0.020}}$   & 11.454  \\
 & 0.6-0.8 & 1019+28 & 0.230$^{{\rm +0.031}}_{{\rm -0.032}}$ & -0.631$^{{\rm +0.127}}_{{\rm -0.123}}$ & 1.601$^{{\rm +0.650}}_{{\rm -0.431}}$ & -0.623$^{{\rm +0.020}}_{{\rm -0.022}}$ & -0.490$^{{\rm +0.015}}_{{\rm -0.015}}$   & 25.423  \\
 & 0.8-1.0 & 1466+23 & 0.223$^{{\rm +0.027}}_{{\rm -0.028}}$ & -0.436$^{{\rm +0.116}}_{{\rm -0.113}}$ & 2.008$^{{\rm +0.704}}_{{\rm -0.487}}$ & -0.455$^{{\rm +0.020}}_{{\rm -0.022}}$ & -0.324$^{{\rm +0.014}}_{{\rm -0.015}}$   & 38.968  \\
 & 1.0-1.2 & 947 +31 & 0.280$^{{\rm +0.031}}_{{\rm -0.032}}$ & -0.496$^{{\rm +0.131}}_{{\rm -0.124}}$ & 0.699$^{{\rm +0.252}}_{{\rm -0.170}}$ & -0.323$^{{\rm +0.028}}_{{\rm -0.028}}$ & -0.167$^{{\rm +0.020}}_{{\rm -0.020}}$   & 23.345  \\
 & 1.2-1.4 & 885 +17 & 0.281$^{{\rm +0.034}}_{{\rm -0.035}}$ & -0.345$^{{\rm +0.137}}_{{\rm -0.132}}$ & 0.639$^{{\rm +0.255}}_{{\rm -0.171}}$ & -0.169$^{{\rm +0.030}}_{{\rm -0.028}}$ & -0.013$^{{\rm +0.020}}_{{\rm -0.021}}$   & 19.457  \\

\hline
11-11.5  
 & combined &  & 0.420$^{{\rm +0.029}}_{{\rm -0.029}}$ & -2.565$^{{\rm +0.110}}_{{\rm -0.114}}$ & 0.419$^{{\rm +0.121}}_{{\rm -0.094}}$ & -1.933$^{{\rm +0.028}}_{{\rm -0.028}}$ & -1.691$^{{\rm +0.026}}_{{\rm -0.026}}$   & 51.328  \\
 & 0.2-0.4 & 100+0 & 0.481$^{{\rm +0.130}}_{{\rm -0.118}}$ & -2.297$^{{\rm +0.414}}_{{\rm -0.530}}$ & 0.026$^{{\rm +0.046}}_{{\rm -0.029}}$ & -1.455$^{{\rm +0.128}}_{{\rm -0.134}}$ & -1.161$^{{\rm +0.194}}_{{\rm -0.154}}$   & 16.792  \\
 & 0.4-0.6 & 132+0 & 0.425$^{{\rm +0.102}}_{{\rm -0.101}}$ & -1.894$^{{\rm +0.349}}_{{\rm -0.387}}$ & 0.040$^{{\rm +0.057}}_{{\rm -0.024}}$ & -1.245$^{{\rm +0.078}}_{{\rm -0.086}}$ & -1.000$^{{\rm +0.117}}_{{\rm -0.093}}$   & 6.035  \\
 & 0.6-0.8 & 200+0 & 0.359$^{{\rm +0.073}}_{{\rm -0.075}}$ & -1.508$^{{\rm +0.269}}_{{\rm -0.276}}$ & 0.077$^{{\rm +0.075}}_{{\rm -0.036}}$ & -1.077$^{{\rm +0.056}}_{{\rm -0.062}}$ & -0.878$^{{\rm +0.056}}_{{\rm -0.053}}$   & 2.084  \\
 & 0.8-1.0 & 315+0 & 0.385$^{{\rm +0.060}}_{{\rm -0.062}}$ & -1.370$^{{\rm +0.232}}_{{\rm -0.239}}$ & 0.073$^{{\rm +0.052}}_{{\rm -0.029}}$ & -0.857$^{{\rm +0.052}}_{{\rm -0.056}}$ & -0.641$^{{\rm +0.043}}_{{\rm -0.043}}$   & 2.618  \\
 & 1.0-1.2 & 199+0 & 0.438$^{{\rm +0.084}}_{{\rm -0.088}}$ & -1.420$^{{\rm +0.358}}_{{\rm -0.360}}$ & 0.028$^{{\rm +0.032}}_{{\rm -0.014}}$ & -0.727$^{{\rm +0.084}}_{{\rm -0.092}}$ & -0.471$^{{\rm +0.067}}_{{\rm -0.066}}$   & 7.740  \\
 & 1.2-1.4 & 194+0 & 0.413$^{{\rm +0.087}}_{{\rm -0.103}}$ & -1.116$^{{\rm +0.438}}_{{\rm -0.399}}$ & 0.026$^{{\rm +0.037}}_{{\rm -0.022}}$ & -0.509$^{{\rm +0.114}}_{{\rm -0.114}}$ & -0.272$^{{\rm +0.067}}_{{\rm -0.072}}$   & 10.272  \\

\hline

\end{tabular}
\caption{Best-fit parameters assuming a double exponential profile
  fitted over the 1/V$_{\rm max}$ non-parametric $s\SFR$ functions.
  In the lowest stellar mass bin $9.5<log(\Ms)<10$, we set the value
  of $\sigma$ which is not  constrained. Systematic uncertainties are not
  included. We consider that a systematic uncertainty of $+0.1$ dex
  could affect $\SFR$ measurement (see \S\ref{data}). Assuming a
  systematic uncertainty of $\pm$0.1 dex in the stellar mass, we
  obtain a systematic uncertainty of $^{+0.14}_{-0.1}$ in the
  $log(s\SFR^*)$, the median and the averaged $s\SFR$ estimates. 
  \label{paraDoubleExpo}}
\end{center}
\end{table*}

\begin{table*}[htb!]
\begin{center}
\begin{tabular}{l c c c c c c c c} \hline

         &            &         &                                  &                                 &            &  median  &  average &   \\ 
         &            & N COSMOS        &                                  &  $log(s\SFR^*)$           &   $\Phi^*$  &   $log(s\SFR)$ &   $log(s\SFR)$ & $\chi^2$ \\ 
$\Ms$ bin &    z-bin   &  + GOODS      &    $\sigma$                      &  (Gyr$^{-1}$)              &  ($10^{-3} Mpc^{-3}$)   &  (Gyr$^{-1}$)    &  (Gyr$^{-1}$)    &   \vspace{0.2cm} \\ \hline
\hline
9.5-10 
 & combined &  & 0.280$^{{\rm +0.011}}_{{\rm -0.010}}$ & -1.037$^{{\rm +0.013}}_{{\rm -0.014}}$ & 0.970$^{{\rm +0.010}}_{{\rm -0.010}}$ & -1.027$^{{\rm +0.012}}_{{\rm -0.014}}$ & -0.913$^{{\rm +0.008}}_{{\rm -0.009}}$   & 158.167  \\
 & 0.2-0.4 & 1080+0  & 0.27 & -0.685$^{{\rm +0.012}}_{{\rm -0.012}}$ & 3.209$^{{\rm +0.077}}_{{\rm -0.077}}$ & -0.675$^{{\rm +0.012}}_{{\rm -0.012}}$ & -0.567$^{{\rm +0.012}}_{{\rm -0.012}}$   & 16.010  \\  
 & 0.4-0.6 & 643 +45 & 0.27 & -0.549$^{{\rm +0.013}}_{{\rm -0.014}}$ & 1.912$^{{\rm +0.041}}_{{\rm -0.041}}$ & -0.539$^{{\rm +0.012}}_{{\rm -0.012}}$ & -0.430$^{{\rm +0.013}}_{{\rm -0.013}}$   & 13.148  \\  
 & 0.6-0.8 & 1009+92 & 0.27 & -0.328$^{{\rm +0.010}}_{{\rm -0.010}}$ & 2.296$^{{\rm +0.037}}_{{\rm -0.037}}$ & -0.319$^{{\rm +0.010}}_{{\rm -0.010}}$ & -0.209$^{{\rm +0.010}}_{{\rm -0.010}}$   & 16.896  \\  
 & 0.8-1.0 & 1111+61 & 0.27 & -0.239$^{{\rm +0.009}}_{{\rm -0.010}}$ & 2.362$^{{\rm +0.032}}_{{\rm -0.032}}$ & -0.229$^{{\rm +0.010}}_{{\rm -0.010}}$ & -0.120$^{{\rm +0.009}}_{{\rm -0.009}}$   & 23.470  \\  
 & 1.0-1.2 & 546 +60 & 0.27 & -0.056$^{{\rm +0.015}}_{{\rm -0.016}}$ & 1.614$^{{\rm +0.024}}_{{\rm -0.024}}$ & -0.047$^{{\rm +0.016}}_{{\rm -0.014}}$ & 0.063$^{{\rm +0.015}}_{{\rm -0.015}}$   & 38.485  \\   
 & 1.2-1.4 & 180 +29 & 0.27 & -0.002$^{{\rm +0.022}}_{{\rm -0.026}}$ & 1.596$^{{\rm +0.022}}_{{\rm -0.022}}$ & 0.009$^{{\rm +0.022}}_{{\rm -0.024}}$ & 0.117$^{{\rm +0.023}}_{{\rm -0.023}}$   & 10.858  \\    

\hline
10-10.5  
 & combined &  & 0.307$^{{\rm +0.005}}_{{\rm -0.005}}$ & -1.194$^{{\rm +0.008}}_{{\rm -0.008}}$ & 0.956$^{{\rm +0.010}}_{{\rm -0.010}}$ & -1.183$^{{\rm +0.008}}_{{\rm -0.008}}$ & -1.053$^{{\rm +0.006}}_{{\rm -0.006}}$   & 390.506  \\
 & 0.2-0.4 & 991 +0  & 0.262$^{{\rm +0.013}}_{{\rm -0.012}}$ & -0.736$^{{\rm +0.015}}_{{\rm -0.018}}$ & 1.813$^{{\rm +0.049}}_{{\rm -0.047}}$ & -0.727$^{{\rm +0.014}}_{{\rm -0.018}}$ & -0.622$^{{\rm +0.012}}_{{\rm -0.013}}$   & 43.939  \\
 & 0.4-0.6 & 1193+31 & 0.278$^{{\rm +0.012}}_{{\rm -0.011}}$ & -0.645$^{{\rm +0.020}}_{{\rm -0.021}}$ & 1.249$^{{\rm +0.034}}_{{\rm -0.033}}$ & -0.635$^{{\rm +0.020}}_{{\rm -0.020}}$ & -0.522$^{{\rm +0.015}}_{{\rm -0.016}}$   & 64.749  \\
 & 0.6-0.8 & 1698+44 & 0.302$^{{\rm +0.011}}_{{\rm -0.010}}$ & -0.445$^{{\rm +0.015}}_{{\rm -0.016}}$ & 1.344$^{{\rm +0.028}}_{{\rm -0.028}}$ & -0.433$^{{\rm +0.014}}_{{\rm -0.016}}$ & -0.307$^{{\rm +0.012}}_{{\rm -0.012}}$   & 23.998  \\
 & 0.8-1.0 & 2239+40 & 0.328$^{{\rm +0.010}}_{{\rm -0.010}}$ & -0.256$^{{\rm +0.015}}_{{\rm -0.016}}$ & 1.379$^{{\rm +0.025}}_{{\rm -0.025}}$ & -0.245$^{{\rm +0.016}}_{{\rm -0.014}}$ & -0.102$^{{\rm +0.011}}_{{\rm -0.011}}$   & 60.545  \\
 & 1.0-1.2 & 1570+71 & 0.323$^{{\rm +0.012}}_{{\rm -0.011}}$ & -0.103$^{{\rm +0.017}}_{{\rm -0.017}}$ & 0.951$^{{\rm +0.019}}_{{\rm -0.019}}$ & -0.091$^{{\rm +0.016}}_{{\rm -0.018}}$ & 0.048$^{{\rm +0.012}}_{{\rm -0.013}}$   & 40.119  \\
 & 1.2-1.4 & 828 +34 & 0.298$^{{\rm +0.034}}_{{\rm -0.030}}$ & 0.002$^{{\rm +0.038}}_{{\rm -0.045}}$ & 0.863$^{{\rm +0.017}}_{{\rm -0.017}}$ & 0.013$^{{\rm +0.038}}_{{\rm -0.044}}$ & 0.138$^{{\rm +0.023}}_{{\rm -0.025}}$   & 10.214  \\

\hline
10.5-11  
 & combined& & 0.385$^{{\rm +0.009}}_{{\rm -0.008}}$ & -1.504$^{{\rm +0.013}}_{{\rm -0.013}}$ & 0.948$^{{\rm +0.010}}_{{\rm -0.010}}$ & -1.491$^{{\rm +0.014}}_{{\rm -0.012}}$ & -1.307$^{{\rm +0.008}}_{{\rm -0.008}}$   & 218.529  \\
 & 0.2-0.4 & 539 +0 & 0.379$^{{\rm +0.025}}_{{\rm -0.023}}$ & -1.056$^{{\rm +0.034}}_{{\rm -0.037}}$ & 0.838$^{{\rm +0.034}}_{{\rm -0.034}}$ & -1.041$^{{\rm +0.032}}_{{\rm -0.038}}$ & -0.863$^{{\rm +0.026}}_{{\rm -0.027}}$   & 10.334  \\
 & 0.4-0.6 & 697 +11& 0.367$^{{\rm +0.026}}_{{\rm -0.021}}$ & -0.878$^{{\rm +0.030}}_{{\rm -0.039}}$ & 0.636$^{{\rm +0.024}}_{{\rm -0.022}}$ & -0.865$^{{\rm +0.030}}_{{\rm -0.038}}$ & -0.696$^{{\rm +0.021}}_{{\rm -0.025}}$   & 19.190  \\
 & 0.6-0.8 & 1019+28& 0.362$^{{\rm +0.031}}_{{\rm -0.027}}$ & -0.660$^{{\rm +0.039}}_{{\rm -0.046}}$ & 0.611$^{{\rm +0.023}}_{{\rm -0.022}}$ & -0.647$^{{\rm +0.038}}_{{\rm -0.044}}$ & -0.481$^{{\rm +0.022}}_{{\rm -0.025}}$   & 49.577  \\
 & 0.8-1.0 & 1466+23& 0.384$^{{\rm +0.017}}_{{\rm -0.017}}$ & -0.553$^{{\rm +0.029}}_{{\rm -0.029}}$ & 0.740$^{{\rm +0.019}}_{{\rm -0.019}}$ & -0.539$^{{\rm +0.030}}_{{\rm -0.028}}$ & -0.356$^{{\rm +0.019}}_{{\rm -0.018}}$   & 66.046  \\
 & 1.0-1.2 & 947 +31& 0.379$^{{\rm +0.019}}_{{\rm -0.018}}$ & -0.368$^{{\rm +0.028}}_{{\rm -0.028}}$ & 0.452$^{{\rm +0.013}}_{{\rm -0.013}}$ & -0.355$^{{\rm +0.028}}_{{\rm -0.028}}$ & -0.176$^{{\rm +0.019}}_{{\rm -0.020}}$   & 28.941  \\
 & 1.2-1.4 & 885 +17& 0.371$^{{\rm +0.023}}_{{\rm -0.021}}$ & -0.202$^{{\rm +0.028}}_{{\rm -0.030}}$ & 0.415$^{{\rm +0.012}}_{{\rm -0.012}}$ & -0.187$^{{\rm +0.026}}_{{\rm -0.030}}$ & -0.016$^{{\rm +0.020}}_{{\rm -0.020}}$   & 22.440  \\

\hline
11-11.5  
 & 0.1-0.2 &  & 0.463$^{{\rm +0.024}}_{{\rm -0.022}}$ & -1.953$^{{\rm +0.027}}_{{\rm -0.029}}$ & 0.971$^{{\rm +0.010}}_{{\rm -0.010}}$ & -1.937$^{{\rm +0.028}}_{{\rm -0.028}}$ & -1.686$^{{\rm +0.025}}_{{\rm -0.025}}$   & 50.587  \\
 & 0.2-0.4 & 100+0 & 0.478$^{{\rm +0.129}}_{{\rm -0.099}}$ & -1.450$^{{\rm +0.118}}_{{\rm -0.129}}$ & 0.101$^{{\rm +0.012}}_{{\rm -0.012}}$ & -1.433$^{{\rm +0.118}}_{{\rm -0.126}}$ & -1.167$^{{\rm +0.172}}_{{\rm -0.147}}$   & 16.654  \\
 & 0.4-0.6 & 132+0 & 0.442$^{{\rm +0.086}}_{{\rm -0.070}}$ & -1.243$^{{\rm +0.077}}_{{\rm -0.089}}$ & 0.095$^{{\rm +0.008}}_{{\rm -0.008}}$ & -1.227$^{{\rm +0.078}}_{{\rm -0.088}}$ & -0.995$^{{\rm +0.099}}_{{\rm -0.083}}$   & 5.850  \\
 & 0.6-0.8 & 200+0 & 0.401$^{{\rm +0.057}}_{{\rm -0.049}}$ & -1.076$^{{\rm +0.059}}_{{\rm -0.067}}$ & 0.102$^{{\rm +0.006}}_{{\rm -0.006}}$ & -1.061$^{{\rm +0.058}}_{{\rm -0.066}}$ & -0.865$^{{\rm +0.051}}_{{\rm -0.052}}$   & 1.891  \\
 & 0.8-1.0 & 315+0 & 0.434$^{{\rm +0.051}}_{{\rm -0.045}}$ & -0.871$^{{\rm +0.056}}_{{\rm -0.063}}$ & 0.121$^{{\rm +0.007}}_{{\rm -0.007}}$ & -0.855$^{{\rm +0.054}}_{{\rm -0.062}}$ & -0.631$^{{\rm +0.043}}_{{\rm -0.044}}$   & 2.934  \\
 & 1.0-1.2 & 199+0 & 0.502$^{{\rm +0.075}}_{{\rm -0.063}}$ & -0.779$^{{\rm +0.081}}_{{\rm -0.085}}$ & 0.078$^{{\rm +0.005}}_{{\rm -0.005}}$ & -0.761$^{{\rm +0.080}}_{{\rm -0.084}}$ & -0.472$^{{\rm +0.069}}_{{\rm -0.065}}$   & 7.464  \\
 & 1.2-1.4 & 194+0 & 0.489$^{{\rm +0.070}}_{{\rm -0.061}}$ & -0.570$^{{\rm +0.091}}_{{\rm -0.095}}$ & 0.058$^{{\rm +0.004}}_{{\rm -0.004}}$ & -0.551$^{{\rm +0.088}}_{{\rm -0.094}}$ & -0.276$^{{\rm +0.062}}_{{\rm -0.067}}$   & 9.662  \\

\hline

\end{tabular}
\caption{Best-fit parameters assuming a log-normal function over the
  1/V$_{\rm max}$ non-parametric $s\SFR$ functions. In the lowest
  stellar mass bin $9.5<log(\Ms)<10$, we are not able to constrain the
  value of $\sigma$ which is fixed.  As in Table
  \ref{paraDoubleExpo}, a systematic uncertainty of $^{+0.14}_{-0.1}$
  in the $log(s\SFR^*)$, the median and the averaged $s\SFR$ estimates
  should be added to the given uncertainties.
\label{paraGauss}}
\end{center}
\end{table*}

\begin{table*}[htb!]
\begin{center}
\begin{tabular}{l c c c } \hline
mass bin &  $a$ &   $\beta$  &  $b$  \\
\hline
double-exponential &   & & \\
\hline
all           &  $-1.07\pm0.02$  & $-0.172\pm0.007$ & $3.14\pm0.07$ \\
9.5-10.0      &  $-1.07\pm0.03$  &  & $2.88\pm0.12$ \\
10-10.5       &  $-1.17\pm0.02$  &  & $3.31\pm0.10$ \\
10.5-11       &  $-1.45\pm0.04$  &  & $3.52\pm0.15$ \\
11-11.5       &  $-1.92\pm0.16$  &  & $3.78\pm0.60$ \\
\hline
log-normal &   & & \\
\hline
all         &  $-1.02\pm0.02$  & $-0.201\pm0.008$ & $3.09\pm0.07$  \\
9.5-10.0    &  $-1.01\pm0.02$  &  & $2.88\pm0.12$ \\
10-10.5     &  $-1.12\pm0.02$  &  & $3.12\pm0.10$ \\
10.5-11     &  $-1.46\pm0.04$  &  & $3.44\pm0.16$ \\
11-11.5     &  $-1.85\pm0.15$  &  & $3.50\pm0.50$ \\
\end{tabular}

\caption{Best-fit parameters describing the $s\SFR$ evolution as a function of
  redshift and mass following the parametrization given in
  Eq.\ref{ssfrEvol}.
  \label{parametrisation}}
\end{center}
\end{table*}

\section{Conclusions}\label{conclusions}

We characterize the shape and the evolution of the star-forming main
sequence by measuring the $s\SFR$ function, i.e., the number density in
a comoving volume (in Mpc$^{-3}$) and per logarithmic bin of $s\SFR$
(in dex$^{-1}$) of star-forming galaxies. We combine the data from the
GOODS and the COSMOS surveys and we derive the $s\SFR$ functions at
$0.2<z<1.4$ in four stellar mass bins between $10^{9.5}\Msol < \Ms <
10^{11.5}\Msol$. We show that the GOODS and the COSMOS surveys do not
cover the same area in the $\Ms$-$s\SFR$ plane, which demonstrates the
importance of taking into account selection effects in the study of
the main sequence.

We base our analysis on a MIPS 24 $\mu m$ selected catalogue, adding
Herschel data when possible. We estimate the $\SFR$ by summing the
contribution of the IR and UV light. While our conclusions are based
on the $s\SFR_{UV+IR}$ functions, we also measure the $s\SFR$
functions using optically based tracers of the $\SFR$.

We estimate the $s\SFR$ functions of star-forming galaxies using
several non-parametric estimators. We select the star-forming
population using the presence of a bimodal distribution in the
$M_{NUV}-M_R$/$M_R-M_K$ plane and we check that our conclusions are not
too sensitive to the exact position of the selection criterion. We fit
the non-parametric $s\SFR$ functions by considering two possible
profiles: a log-normal function and a double-exponential function. We
add a starburst component to the $s\SFR$ function and we also add an
additional constraint in the fitting procedure using the $\MF$.

Based on our $s\SFR$ functions, we derive the evolution of the median
and average $s\SFR$. We obtain a clear increase in the $s\SFR$ as a
function of redshift as $\propto (1+z)^b$. Assuming that the $s\SFR$
evolution does not depend on the mass, we find $b=3.18\pm 0.06$. If we
allow $b$ to depend on the mass, we obtain that the evolution is
faster for massive galaxies: $b$ varies from $b=2.88\pm 0.12$ at
$\Ms=10^{9.5}\Msol$ to $b=3.78\pm 0.60$ at $\Ms=10^{11.5}\Msol$. Our
observed evolution of the $s\SFR$ is consistent with the evolution of
the $sMIR_{DM}$ for $\Ms<10^{10}\Msol$ galaxies, but deviates from it
at higher masses.

We also compare our results with the predictions of a semi-analytical
model from Wang et al. (2008). While the predicted $s\SFR$ functions
could be parametrized by a double-exponential profile and matches our
results at $\Ms<10^{10.5}\Msol$ reasonably well, we observe that the
agreement breaks down for massive galaxies at high $s\SFR$. The
description of the recipes impacting the SFH of massive galaxies
should probably be improved in this SAM.

We note that even at $z<1$, it is challenging to constrain the full
shape of the $s\SFR$ functions. Dust-free tracers of the $\SFR$ do not
reach a sufficiently deep $\SFR$ limit to sample well below the peak
in $s\SFR$, while tracers based on the optical are prone to large
biases because of uncertain dust corrections. Still, we combine all
non-parametric estimates of the $s\SFR$ functions at $z=0$. We find
that the shape of the $s\SFR$ distribution seems invariant with time
at $z<1.4$ but depends on the mass. We observe a broadening of the
main sequence with $\Ms$. Assuming a log-normal distribution, we find
that $\sigma$ does not vary with redshift at $z<1.4$, and increases
from 0.28 to 0.46 dex between $log(\Ms)=9.5-10$ and $log(\Ms)=11-11.5$
dex. While the stochasticity of the star formation in individual
galaxies could explain the width of the $s\SFR$ function at low mass,
it cannot explain an increase in this width with $\Ms$. A possibility
is that the SFHs become more diverse as the mass increases, as a
result of the numerous processes that reduce the star formation in
massive galaxies.

We also show that the evolution of the median $s\SFR$ in a logarithmic
scale decreases as $-0.17\Ms$. We note that the commonly adopted
linear relation between $log(s\SFR)$ and $log(\Ms)$ is not suitable
for our data. Such a dependency with $\Ms$ at high mass could be
reproduced by assuming exponentially declining SFH with $\tau$ having
an inverse dependency with mass $\tau \propto 1/\M$. Several processes
could reduce the $s\SFR$ as the stellar mass increases.  Accretion of
cold gas can be suppressed in hot gas halos ($M_{H}>10^{12}\Msol$)
leading to gas exhaustion in the central galaxies. This process should
occur on longer timescales ($>3-4Gyr$) than usually assumed to explain
our observed trend. Another possibility is that the efficiency of the
star formation is declining toward massive sources: disks are more
settled and stabilized against fragmentation as the mass
increases. The presence of a bulge could be crucial in reducing the
star formation efficiency. Finally, a combined analysis of the $s\SFR$
functions and of the quiescent ${\cal GSMF}$ could constrain the
relative importance between secular and quenching processes by setting
the quenching timescale with the ${\cal GSMF}$ evolution.

\begin{acknowledgements}
  We are grateful to the referee for the careful reading of the
  manuscript and his/her useful suggestions. We thank Samuel Boissier,
  V\'eronique Buat, Andrea Cattaneo, Jared Gabor, Sylvain De La Torre
  and Mark Sargent for useful discussions. We thank Carlota Gruppioni
  for providing her data. We gratefully acknowledge the contributions
  of the entire COSMOS collaboration consisting of more than 100
  scientists.  The {\it HST} COSMOS program was supported through NASA
  grant HST-GO-09822.  More information on the COSMOS survey is
  available at http://www.astro.caltech.edu/cosmos. This paper is
  based on observations made with ESO Telescopes at the La Silla
  Paranal Observatory under ESO programme ID 179.A-2005 and on data
  products produced by TERAPIX and the Cambridge Astronomy Survey Unit
  on behalf of the UltraVISTA consortium. LMD acknowledges support
  from the Lyon Institute of Origins under grant ANR-10-LABX-66. AK
  acknowledges support by the Collaborative Research Council 956,
  sub-project A1, funded by the Deutsche Forschungsgemeinschaft (DFG).
\end{acknowledgements}

\appendix

\section{Comparison between $s\SFR$ and $sMIR_{DM}$: impact of the mass loss}\label{loss}

If the feeding in gas is driven by the growth of the dark matter halos
and if the fraction of the infalling gas converted into stars stays
constant, we expect that $\SFR/\int^t_0
SFR(t')dt'=\dot{M}_{H}/M_H=sMIR_{DM}$. However, only a fraction of the
mass created stays trapped in an old stellar population. Therefore,
$\SFR/\int^t_0 SFR(t')dt'$ is different from the $s\SFR$ defined as
$\SFR/\int^t_0 SFR(t')(1-R[t-t'])dt'$ with $R$ the return fraction
depending on the age of the stellar populations (Renzini A. \& Buzzoni
A., 1986).

Assuming a constant $R$ value, we expect $s\SFR'=sMIR_{DM}/(1-R)$. For
the Chabrier (2003) IMF, the maximum value of $R$ is 0.5. Therefore,
we use this value to define the upper bound of the green area in
Fig.\ref{sSFRweinmann} and Fig.\ref{sSFRsimu}.

\begin{figure}[htb!]
\centering \includegraphics[width=9.cm]{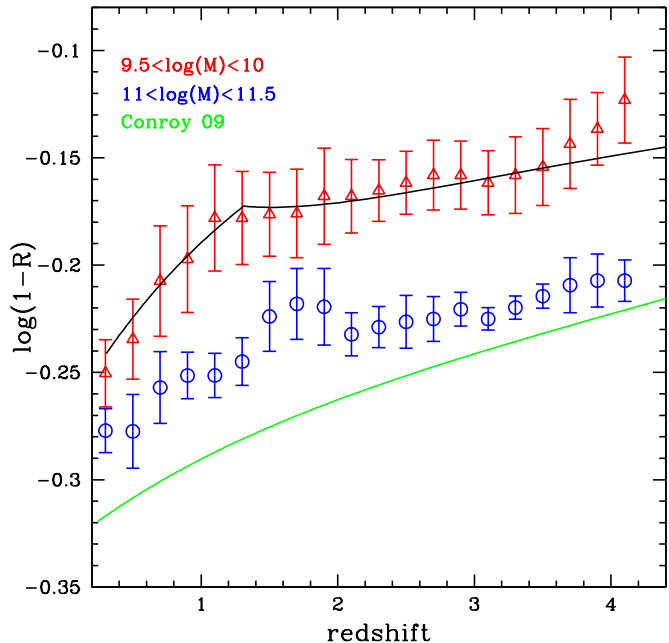}
\caption{Open triangles and open circles represent the median $(1-R)$
  obtained for a low-mass galaxy sample ($9.5<log(\Ms)<10$) and
  high-mass sample ($11<log(\Ms)<11.5$), respectively. The green
  curves correspond to the mass loss parametrized by Conroy \&
  Wechsler (2009) assuming the same redshift of formation $z=10$ for
  all the stars. The black line is the parametrization that we adopt
  as a lower limit for the return fraction.}
           \label{massloss} 
\end{figure}

However, $R$ depends on time. The value of the stellar mass lost by a
given galaxy depends on its SFH. In order to determine the mass lost
along the galaxy history, for each galaxy in our sample we measure the
difference between the total mass obtained by integrating the SFH
(without taking into account the stellar mass loss) and the stellar
mass. Then, we measure the median of these differences as a function
of redshift. Figure \ref{massloss} shows the median stellar mass
effectively lost as a function of redshift $R_{med}(z)$. The lower
bound of the green shaded area in Fig.\ref{sSFRweinmann} and
Fig.\ref{sSFRsimu} corresponds to $sMIR_{DM}/(1-R_{med}(z))$ measured
for low-mass galaxies. We check that the averaged stellar mass
effectively lost in the higher mass bin would be within our two
boundaries.

We note that the evolution of $\frac{sMIR_{DM}}{1-R_{med}(z)}$ is
flatter than $sMIR_{DM}/2$ (i.e., taking $R$ as a constant), since
$R_{med}(z)$ is smaller at high redshift than at low redshift (stellar
populations are younger).

\section{Additional selection effects linked to the broadening of the $\Ms-\SFR$ relation}\label{broadsel}

In this appendix, we investigate wether the criterion used to select
star-forming galaxies could artificially create a broadening of the
$s\SFR$ function with the mass.

We first check that the intrinsic evolution of the $s\SFR$ does not
enlarge our estimate of $\sigma$ significantly (e.g., Speagle et
al. 2014). Using a simple model with $s\SFR \propto (1+z)^{3.8}$ (our
extreme value of $b$), we find that the broadening cannot be
overestimated by more than 0.02 dex in our redshift bins owing to the
$s\SFR$ intrinsic evolution.

\begin{figure}[htb!]
\centering \includegraphics[width=9.cm]{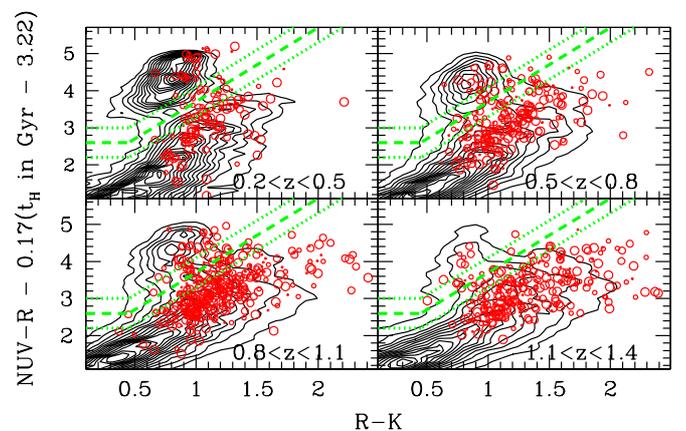}
\caption{Same as Fig.\ref{NRK}, except that the red circles are the
  massive 24$\mu m$ sources ($11<log(\Ms)<11.5$) and the size of the
  sources is proportional to the 24$\mu m$ flux. The contours refer to
  the full galaxy sample at $log(\Ms)>9.5$. The largest fraction of
  massive galaxies are well below the selection criterion and the
  brightest ones are located in top right part of the diagram with the
  most extinguished sources.}
           \label{NRKmassive} 
\end{figure}

Since the rest-frame colors are closely correlated with the $s\SFR$,
one could artificially modify the shape of the $s\SFR$ function
depending on the rest-frame color cut used to separate quiescent and
star-forming galaxies. In particular, it is unclear if the population
that is transitioning from the star-forming to the quiescent
population should be included in the analysis, and how it affects the
$s\SFR$ function. As shown in Fig.\ref{NRKmassive}, the most massive
galaxies tend to lie much closer to the transitioning area than the
other star-forming galaxies. This result is not surprising since Peng
et al. (2010) show that the probability of a star-forming galaxy being
quenched is proportional to its $\SFR$ and mass (their Eq.17). In
order to investigate the impact of the selection criterion, we move
down and up the selection criterion by 0.3 mag which are the extreme
values we could adopt (dotted lines in Fig.\ref{NRKmassive}). We find
that this change has no impact on the $s\SFR$ functions of galaxies
less massive than $\Ms<10^{11}\Msol$ and the value of $\sigma$ remains
above 0.4 dex at $\Ms>10^{11}\Msol$. While the measurements are
slightly modified for the most massive galaxies, the overall shape of
the $s\SFR$ functions is not affected. Therefore, our conclusions are
not sensitive to the adopted limit used to select star-forming galaxies.

We also investigate if the uncertainties associated with the stellar
mass could artificially create such broadening of the $s\SFR$
function. Indeed, uncertainties in the stellar mass could move the
galaxies from one mass bin to another. Since the $s\SFR$ depends on
the mass, it could artificially broaden the $s\SFR$ distribution. The
galaxies at $\Ms<10^{11}\Msol$ could contaminate the most massive
galaxies and artificially add galaxies with a larger $s\SFR$ since the
$s\SFR$ decreases with the stellar mass. In order to test this effect,
we use the semi-analytical model described in \S\ref{SAM}. We add
random errors using a Gaussian distribution having a standard
deviation of 0.1 dex (already larger than our expected uncertainties)
to the predicted $\SFR$ and to the predicted mass. We find that the
$s\SFR$ functions predicted by the model are almost unmodified. If we
adopt uncertainties of 0.2 dex in mass and in $\SFR$, we get
unrealistic predictions at all masses. We note that systematic
uncertainties are not discussed here since they shift the full
distribution.

\begin{thebibliography}{}


\bibitem[{Abramson et al. }]{Abramson14} Abramson L.E., Kelson D.D., Dressler A. et al., 2014, ApJL, 785, L36 
\bibitem[{Arnouts et al.}]{Arnouts02} Arnouts S., Moscardini L., Vanzella E. et al., 2002, MNRAS, 329, 355 
\bibitem[{Arnouts et al. }]{Arnouts07} Arnouts S., Walcher C.J., Le F\`evre O. et al., 2007, A\&A, 476, 137 
\bibitem[{Arnouts et al.}]{Arnouts13} Arnouts S., Le Floc'h E., Chevallard J. et al., 2013, A\&A, 558, A67 
\bibitem[{Baldry et al.}]{Baldry12} Baldry I.K., Driver S.P., Loveday J. et al., 2012, MNRAS, 421, 621 
\bibitem[{Bauer et al.}]{Bauer13} Bauer A.E., Hopkins A.M., Gunawardhana M. et al., 2013, MNRAS, 434, 209 
\bibitem[{Behroozi et al.}]{Behroozi13} Behroozi P.S., Wechsler R.H. \& Conroy C., 2013, 770, 57
\bibitem[{Bell et al. }]{Bell04} Bell E.F., Wolf C., Meisenheimer K. et al., 2004, ApJ, 608, 752 
\bibitem[Bernhard et al]{Bernhard14} Bernhard E., B{\'e}thermin M., Sargent M. et al., 2014, MNRAS, 442, 509 
\bibitem[{Bertin \& Arnouts }]{Bertin96} Bertin E. \& Arnouts S., 1996, A\&AS, 117, 393 
\bibitem[B{\'e}thermin et al.]{Bethermin12} B{\'e}thermin M., Daddi E., Magdis G. et al., 2012, ApJL, 757, L23 
\bibitem[{Boissier et al.}]{Boissier10} Boissier S., Buat V., Ilbert O., 2010, A\&A, 522, A18
\bibitem[{Bouch\'e et al.}]{Bouche10} Bouch\'e N., Dekel A., Genzel R. et al., 2010, ApJ, 718, 1001 
\bibitem[{Bournaud et al.}]{Bournaud07} Bournaud F., Elmegreen B.G., Elmegreen D.M., 2007, ApJ, 670, 237
\bibitem[Bournaud et al.]{Bournaud11} Bournaud F., Chapon D., Teyssier R. et al., 2011, ApJ, 730, 4 
\bibitem[{Brammer et al. }]{Brammer11} Brammer G.B., Whitaker K.E., van Dokkum P.G. et al., 2011, ApJ, 739, 24 
\bibitem[{Brusa et al. }]{Brusa07} Brusa M., Zamorani G., Comastri A. et al., 2007, ApJS, 172, 353
\bibitem[{Bruzual \& Charlot}]{Bruzual03} Bruzual G. \& Charlot S., 2003, MNRAS, 344, 1000 
\bibitem[{Calzetti et al. }]{Calzetti00} Calzetti D., Armus L., Bohlin R.C. et al., 2000, ApJ, 533, 682
\bibitem[{Capak et al. }]{Capak07} Capak P., Abraham R.G., Ellis R.S. et al., 2007, ApJS, 172, 284 
\bibitem[{Cattaneo et al.}]{Cattaneo06} Cattaneo A., Dekel A., Devriendt J., Guiderdoni B., \& Blaizot J., 2006, MNRAS, 370, 1651 
\bibitem[{Chabrier }]{Chabrier03} Chabrier G., 2003, PASP, 115, 763
\bibitem[Chary \& Elbaz]{Chary01} Chary R. \& Elbaz D., 2001, ApJ, 556, 562 
\bibitem[{Cirasuolo et al. }]{Cirasuolo07} Cirasuolo M., McLure R.J., Dunlop J.S. et al., 2007, MNRAS, 380, 585 
\bibitem[{Comparat et al. }]{Comparat15} Comparat J., Richard J.,  Kneib J.P. et al., 2015, A\&A, 575, AA40 
\bibitem[Conroy et al.]{Conroy09} Conroy C. \& Wechsler R. H., 2009, ApJ, 696, 620
\bibitem[Coupon et al.]{Coupon14} Coupon J., Arnouts S., van Waerbeke L. et al., MNRAS, accepted, astro-ph/1502.02867
\bibitem[Cowie et al.(1996)]{Cowie1996} Cowie L.L., Songaila A., Hu E.M. \& Cohen J.G., 1996, AJ, 112, 839 
\bibitem[{Croton et al. }]{Croton06} Croton D.J., Springel V., White S.D.M. et al., 2006, MNRAS, 365, 11 
\bibitem[{Daddi et al}]{Daddi07} Daddi E., Dickinson M., Morrison G. et al., 2007, ApJ, 670, 156 
\bibitem[{Dale \& Helou }]{Dale02} Dale D.A. \& Helou G. 2002, ApJ, 576, 159 
\bibitem[Dekel et al.(2009)]{2009Natur.457..451D} Dekel A., Birnboim Y., Engel G. et al., 2009, Nature, 457, 451 
\bibitem[De Lucia \& Blaizot]{delucia07} De Lucia G. \& Blaizot J., 2007, MNRAS, 375, 2 
\bibitem[Dey et al.]{Dey08} Dey A., Soifer B.T., Desai V. et al., 2008, ApJ, 677, 943 
\bibitem[Dom{\'{\i}}nguez et al.]{Dominguez2014} Dom{\'{\i}}nguez A.,  Siana B.,Brooks A.M. et al., 2014, astro-ph/1408.5788
\bibitem[Donley et al.]{Donley12} Donley J.L., Koekemoer A.~M., Brusa M. et al., 2012, ApJ, 748, 142 
\bibitem[{Elbaz et al.}]{Elbaz07} Elbaz D., Daddi E., Le Borgne D. et al., 2007, A\&A, 468, 33 
\bibitem[Elbaz et  al.]{Elbaz10} Elbaz D., Hwang H.S., Magnelli B. et al., 2010, A\&A, 518, L29 
\bibitem[Elbaz et  al.]{Elbaz11} Elbaz D., Dickinson M., Hwang H.S. et al., 2011, A\&A, 533, A119 
\bibitem[{Efstathiou et al. }{1988}]{Efstathiou88} Efstathiou G., Ellis R.S., Peterson B.A., 1988, MNRAS, 232, 431 
\bibitem[Faber al.]{Faber} Faber S.M., Willmer C.N.A., Wolf C. et al., 2007, ApJ, 665, 265 
\bibitem[F{\"o}rster-Schreiber et al.]{Fortser14} F{\"o}rster-Schreiber N.M., Genzel R., Newman S.F. et al., 2014, ApJ, 787, 38 
\bibitem[Franzetti et al. ]{Franzetti07} Franzetti P., Scodeggio M., Garilli B. et al., 2007, A\&A, 465, 711 
\bibitem[Fritz et al.]{Fritz14} Fritz A., Scodeggio M., Ilbert O. et al., 2014, A\&A, 563, A92 
\bibitem[Gabor]{Gabor14} Gabor J.M. \& Dav{\'e} R., 2015, MNRAS, 447, 374
\bibitem[Genzel et al.]{Genzel08} Genzel R., Burkert A., Bouch{\'e}, N. et al., 2008, ApJ, 687, 59 
\bibitem[Genzel et al.]{Genzel14} Genzel R., F{\"o}rster Schreiber N.M., Lang P. et al., 2014, ApJ, 785, 75 
\bibitem[Giavalisco et al. ]{Giavalisco04} Giavalisco M., Ferguson H.C., Koekemoer A.M. et al., 2004, ApJL, 600, L93 
\bibitem[Gruppioni et al.]{ Gruppioni13} Gruppioni C., Pozzi F., Rodighiero, G. et al., 2013, MNRAS, 432, 23 
\bibitem[{Guo et al. }]{Guo11} Guo Q., White S., Boylan-Kolchin M. et al., 2011, MNRAS, 413, 101 
\bibitem[Guo et al.]{Guo13} Guo K., Zheng X.Z., \& Fu H., 2013, ApJ, 778, 23 
\bibitem[{Hayward }]{Hayward } Hayward C.C., Lanz L., Ashby M.L.N. et al., 2014, MNRAS, 445, 1598
\bibitem[Heinis et al.]{Heinis13} Heinis S., Buat V., B{\'e}thermin M. et al., 2013, MNRAS, 429, 1113 
\bibitem[Hopkins et al.]{Hopkins08} Hopkins P.F., Hernquist L., Cox T.J., \& Kere{\v s} D., 2008, ApJS, 175, 356 
\bibitem[Hopkins et al.]{Hopkins14} Hopkins P.F., Keres D., Onorbe J. et al., 2014, MNRAS, 445, 581
\bibitem[{Ilbert et al. }]{Ilbert04} Ilbert O., Tresse L., Arnouts S. et al., 2004, MNRAS, 351, 541
\bibitem[{Ilbert et al. }]{Ilbert05} Ilbert O., Tresse L., Zucca E. et al., 2005, A\&A, 439, 863
\bibitem[{Ilbert et al. }]{Ilbert06} Ilbert O., Arnouts S., McCracken H.J. et al., 2006, A\&A, 457, 841 
\bibitem[{Ilbert et al. }]{Ilbert09} Ilbert O., Capak P., Salvato M. et al., 2009, ApJ, 690, 1236 
\bibitem[{Ilbert et al. }]{Ilbert10} Ilbert O., Salvato M., Le Floc'h E. et al., 2010, ApJ, 709, 644 
\bibitem[{Ilbert et al. }]{Ilbert13} Ilbert O., McCracken H., Le F\`evre O. et al., 2013, A\&A, 556, 55
\bibitem[Immeli et al.]{Imeli04} Immeli A., Samland M., Gerhard O., Westera P., 2004, A\&A, 413, 547
\bibitem[Kajisawa et al. ]{Kajisawa10} Kajisawa M., Ichikawa T., Yamada T. et al., 2010, ApJ, 723, 129
\bibitem[Karim et al.]{Karim11} Karim A., Schinnerer E., Mart{\'{\i}}nez-Sansigre A. et al., 2011, ApJ, 730, 61 
\bibitem[Kassin et al.]{Kassin12} Kassin S.A., Weiner B.J., Faber S.M. et al., 2012, ApJ, 758, 106 
\bibitem[Kauffmann et al.]{Kauffmann2000} Kauffmann G. \& Haehnelt M., 2000, MNRAS, 311, 576 
\bibitem[Kelson14]{Kelson14}Kelson D.D., submitted to ApJ, astro-ph/1406.5191
\bibitem[Kennicutt]{Kennicutt98}Kennicutt R.C., 1998, ARA\&A, 36, 189 
\bibitem[Lagache et al.]{Lagache04} Lagache G., Dole H., Puget J.-L. et al., 2004, ApJS, 154, 112 
\bibitem[Lang et al.]{lang14}Lang P., Wuyts S., Somerville R. et al., 2014, ApJ, 788, 11
\bibitem[Lee et al.]{Lee14} Lee N., et al., 2015, ApJ, 801, 80
\bibitem[Le Fevre et al.]{LeFevre04} Le F{\`e}vre O., Vettolani G., Paltani S. et al., 2004, A\&A, 428, 1043 
\bibitem[Le Fevre et al.]{LeFevre14} Le F{\`e}vre O., Amorin R., Bardelli S. et al., 2014, The Messenger, 155, 38 
\bibitem[Le Floc'h et al.]{LeFloch05} Le Floc'h E., Papovich C., Dole H. et al., 2005, ApJ, 632, 169 
\bibitem[Le Floc'h et al.]{LeFloch09} Le Floc'h E., Aussel H., Ilbert O. et al., 2009, ApJ, 703, 222 
\bibitem[{Lynden-Bell }]{Lynden-Bell71} Lynden-Bell D., 1971, MNRAS, 155, 95
\bibitem[{Lilly et al. }]{Lilly07} Lilly S.J., Le F\`evre O., Renzini A. et al., 2007, ApJS, 172, 70
\bibitem[Lilly et al.]{Lilly13} Lilly S.J., Carollo C.M., Pipino A., Renzini A. \& Peng Y., 2013, ApJ, 772, 119 
\bibitem[Lutz et al.]{Lutz11} Lutz D., Poglitsch A., Altieri B. et al., 2011, A\&A, 532, A90 
\bibitem[Magdis et al.]{Magdis12} Magdis G.E., Daddi E., B{\'e}thermin M. et al., 2012, ApJ, 760, 6 
\bibitem[Magnelli et al.]{Magnelli09} Magnelli B., Elbaz D., Chary R.R. et al. 2009, A\&A, 496, 57 
\bibitem[Magnelli et al.]{Magnelli11} Magnelli B., Elbaz D., Chary R.R. et al. 2011, A\&A, 528, A35 
\bibitem[Magnelli et al.]{Magnelli13} Magnelli B., Popesso P., Berta S. et al. 2013, A\&A, 553, A132 
\bibitem[Martig et al.]{Martig09} Martig M., Bournaud F., Teyssier R. \& Dekel A., 2009, ApJ, 707, 250 
\bibitem[McCracken et al.]{McCracken2012} McCracken H.J., Milvang-Jensen B., Dunlop J. et al., 2012, A\&A, 544, 156 
\bibitem[Mignoli et al.]{Mignoli05} Mignoli M., Cimatti A., Zamorani G. et al., 2005, A\&A, 437, 883 
\bibitem[Mitchell et al.]{Mitchell13} Mitchell P.D., Lacey C.G., Baugh C.M., \& Cole S., 2013, MNRAS, 435, 87 
\bibitem[Moster et al.]{Moster2011} Moster B.P., Somerville R.S., Newman J.A., and Rix H., 2011, ApJ, 731, 113
\bibitem[Moustakas et al.]{Moustakas13} Moustakas J., Coil A., Aird J. et al., 2013, ApJ, 767, 50
\bibitem[Muzzin et al.]{Muzzin13} Muzzin A., Marchesini D., Stefanon M. et al., 2013a, ApJS, 206, 8 
\bibitem[Neistein \& Dekel]{Neistein08} Neistein E., \& Dekel A., 2008, MNRAS, 388, 1792 
\bibitem[{Noeske et al.}]{Noeske07} Noeske K.G., Weiner B.J., Faber S.M. et al., 2007a, ApJL, 660, L43 
\bibitem[Noeske et al.]{Noeske07} Noeske K.G., Faber S.M., Weiner B.J. et al., 2007b, ApJL, 660, L47 
\bibitem[{Oke}]{Oke74} Oke J.B., 1974, ApJS, 27, 21 
\bibitem[Oliver et al. ]{Oliver12} Oliver S.J., Bock J., Altieri B. et al., 2012, MNRAS, 424, 1614 
\bibitem[Pannella et al.]{Pannella09} Pannella M., Carilli C.L., Daddi E. et al., 2009, ApJL, 698, L116 
\bibitem[{Peng et al. }]{Peng10} Peng Y.J., Lilly S.J., Kova{\v c} K. et al., 2010, ApJ, 721, 193 
\bibitem[{Peng et al. }]{Peng14} Peng Y.J., Lilly S.J., Renzini A., Carollo M., 2014, ApJ, 790, 95
\bibitem[Perez et al.]{Perez13}Perez J., Valenzuela O., Tissera P.B. \& Michel-Dansac L., 2013, MNRAS, 436, 259
\bibitem[{Pozzetti et al. }]{Pozzetti10} Pozzetti L., Bolzonella M., Zucca E. et al., 2010, A\&A, 523, 13
\bibitem[Rieke et al.]{Rieke09} Rieke G.H., Alonso-Herrero A., Weiner B.J. et al., 2009, ApJ, 692, 556 
\bibitem[{Renzini et al.}]{Renzini86} Renzini A. \& Buzzoni A., 1986, Spectral Evolution of Galaxies, 122, 195 
\bibitem[{Riguccini et al.}]{Riguccini11} Riguccini L., Le Floc'h E., Ilbert O. et al., 2011, A\&A, 534, AA81 
\bibitem[Robitaille \& Whitney]{Robitaille10} Robitaille T.P. \& Whitney B.A., 2010, ApJL, 710, L11 
\bibitem[Rodighiero et al.]{Rodighiero10} Rodighiero G., Vaccari M., Franceschini A. et al., 2010, A\&A, 515, A8 
\bibitem[{Rodighiero et al. }]{Rodighiero11} Rodighiero G., Daddi E., Baronchelli I. et al., 2011, ApJL, 739, L40 
\bibitem[{Rodighiero et al. }]{Rodighiero14}Rodighiero G., Renzini A.,  Daddi E. et al., 2014, MNRAS, 443, 19
\bibitem[Saintonge et al.]{Saintonge12} Saintonge A., Tacconi L.J., Fabello S. et al., 2012, ApJ, 758, 73 
\bibitem[Salim et al.]{Salim07} Salim S., Rich R.M., Charlot S. et al., 2007, ApJS, 173, 267 
\bibitem[Salim  \& Lee]{Salim12} Salim S., \& Lee J.C., 2012, ApJ, 758, 134 
\bibitem[{Salmi et al. }]{Salmi13} Salmi F., Daddi E., Elbaz D. et al., 2012, ApJL, 754, L14 
\bibitem[{Sandage et al. }{1979}]{Sandage79} Sandage A., Tammann G.A., Yahil A., 1979, ApJ, 232, 352 
\bibitem[{Sanders et al. }]{Sanders07} Sanders D.B., Salvato M., Aussel H. et al., 2007, ApJS, 172, 86 
\bibitem[Sargent et al.]{Sargent12} Sargent M.T., B{\'e}thermin M., Daddi E. \& Elbaz D., 2012, ApJL, 747, L31 
\bibitem[{Saunders et al. }]{Saunders90} Saunders W., Rowan-Robinson M., Lawrence A. et al., 1990, MNRAS, 242, 318
\bibitem[{Scoville et al. }]{Scoville07} Scoville N., Aussel H., Brusa M. et al. 2007, ApJS, 172, 1 
\bibitem[{Schmidt }{1968}]{Schmidt68} Schmidt M., 1968, ApJ, 151, 393 
\bibitem[Sheth et al.]{sheth08} Sheth K., Elmegreen D.M., Elmegreen M. et al., 2008, ApJ, 675, 1141
\bibitem[Smol{\v c}i{\'c} et al.]{smolcic08} Smol{\v c}i{\'c} V., Schinnerer E., Scodeggio M. et al., 2008, ApJS, 177, 14 
\bibitem[Snaith et al.]{Snaith14} Snaith O.N., Haywood M., Di Matteo P. et al., 2014, ApJL, 781, L31 
\bibitem[Sparre et al. ]{Sparre14} Sparre M., Hayward C.C., Springel V. et al., 2014, astro-ph/1409.0009 
\bibitem[Speagle et al.]{Speagle14}Speagle J.S., Steinhardt C., Capak P., Silverman J.D., 2014, ApJS, 214, 15
\bibitem[{Tomczak et al. }]{Tomczak14} Tomczak A.R., Quadri R.F., Tran K.-V.H. et al., 2014, ApJ, 783, 85 
\bibitem[Utomo et al.]{Utomo14} Utomo D., Kriek M., Labb{\'e} I., Conroy C. \& Fumagalli M., 2014, ApJL, 783, L30 
\bibitem[van Dokkum et al.]{vanDokkum13} van Dokkum P.G., Leja J., Nelson E.J. et al., 2013, ApJL, 771, L35 
\bibitem[{Wang et al. }]{Wang08} Wang J., De Lucia G., Kitzbichler  M.G. and White S.D.M., 2008, MNRAS, 384, 1301
\bibitem[Weinmann et al.]{Weinmann11} Weinmann S.M., Neistein E. \& Dekel A., 2011, MNRAS, 417, 2737 
\bibitem[Weinmann et al.]{Weinmann12} Weinmann S.M., Pasquali A., Oppenheimer B.D. et al., 2012, MNRAS, 426, 2797 
\bibitem[Whitaker et al.]{Whitaker12} Whitaker K.E., van Dokkum P.G., Brammer G. \& Franx M., 2012, ApJL, 754, L29 
\bibitem[{Whitaker et al.}]{Whitaker14} Whitaker K.E., Franx M., Leja J. et al., 2014, 795, 104
\bibitem[{Williams et al. 2009}]{Williams09} Williams R.J., Quadri R.F., Franx M. et al., 2009, ApJ, 691, 1879
\bibitem[{Wuyts et al. }]{Wuyts08} Wuyts S., Labb{\'e} I., Schreiber N.~M.~F. et al., 2008, ApJ, 682, 985 


\end{thebibliography}
\end{document}